# Reviewing two decades of Energy system analysis with bibliometrics

Dominik Franjo Dominković*, dodo@dtu.dk, Department of Applied Mathematics and Computer Science, Technical University of Denmark, Matematiktorvet, 2800 Kgs. Lyngby, Denmark.

Jann Michael Weinand, jann.weinand@kit.edu, Chair of Energy Economics, Institute for Industrial Production (IIP), Karlsruhe Institute of Technology, Karlsruhe, Germany

Fabian Scheller, fjosc@dtu.dk, Energy Systems Analysis, Division of Sustainability, Department of Technology, Management and Economics, Technical University of Denmark, Matematiktorvet, 2800 Kgs. Lyngby, Denmark.

Matteo D'Andrea, matteo.d'andrea@abdn.ac.uk, Chair of Energy Transition, School of Engineering, University of Aberdeen, King's College, Aberdeen AB24 3FX, United Kingdom

Russell McKenna, russell.mckenna@abdn.ac.uk, Chair of Energy Transition, School of Engineering, University of Aberdeen, King's College, Aberdeen AB24 3FX, United Kingdom

* Corresponding author

## Abstract

The field of Energy System Analysis (ESA) has experienced exponential growth in the number of publications since at least the year 2000. This paper presents a comprehensive bibliometric analysis on ESA by employing different algorithms in Matlab and R. The focus of results is on quantitative indicators relating to number and type of publication outputs, collaboration links between institutions, authors and countries, and dynamic trends within the field. The five and twelve most productive countries have 50% and 80% of ESA publications respectively. The dominant institutions are even more concentrated within a small number of countries. A significant concentration of published papers within countries and institutions was also confirmed by analysing collaboration networks. These show dominant collaboration within the same university or at least the same country. There is also is a strong link among the most successful journals, authors and institutions. The *Energy* journal has had the most publications in the field, and its editor-in-chief Lund H is the author with most of the publications in the field, as well as the author with most of the highly cited publications in the field. In terms of the dynamics within the field in the past decade, recent years have seen a higher impact of topics related to flexibility and hybrid/integrated energy systems alongside a decline in individual technologies. This paper provides a holistic overview of two decades' research output and enables interested readers to obtain a comprehensive overview of the key trends in this active field.

**Highlights:**

- a review of more than 12 000 papers in the field of Energy system analysis

- five most productive countries have 50% of publications in the field

- there is also is a strong link among the most successful journals, authors and institutions

- collaboration networks predominantly include only the institutions within the same country

## Contents

# 1. Introduction

Energy system analysis (ESA) is a very broad interdisciplinary field that has emerged as a research field in its own right in the past three decades. The subfield of energy analysis developed methods to understand energy transformation and use process, mainly motivated by the two oil price hikes of the 1970s and led by proponents Chapman (1974; 1975), and Boustead and Hancock (1979). The energy analysis methods reflected thermodynamic properties such as energy, enthalpy and exergy, and economic ones such as economic output and/or prices. They were subsequently adopted alongside the ideas of systems analysis, which came from defence and aerospace contexts, to embody the area of energy systems analysis. The latter field often adopts theoretical and methodological frameworks from operational research to support more efficient and effective decision-making. Hand-in-hand with rapid developments in computing power since the 1970s, increasingly detailed and complex models have been developed over the ensuing decades – starting in the 1990s, as documented by much of the literature referenced in this article. For example, in the United Kingdom context, almost 100 different models have been developed and/or applied in the past two decades (Hall and Buckley 2016).

More recently, rapid transformations of energy systems, partly inspired by Kyoto and subsequent international agreements and natural disasters (e.g. Fukushima), have provided renewed interest and challenges for this field. The global expansion in renewable energy generation has required more detailed models of technical systems and markets and policies (Pfenninger et al. 2014). Models have had to be enhanced to consider a higher spatial and temporal resolution, new types of supply and increasingly flexible demand. At the same time, the field has also attracted much criticism, mainly directed at the lack of transparency in modelling exercises, for example about assumptions, data sources, and uncertainties, which also relates to communication and interaction with non-expert stakeholders (Strachan et al. 2016; DeCarolis et al. 2017).

Indeed, the strong interest in ESA in recent decades has inspired many reviews, some of which have common ground with the present paper[1]. While different narrative or systematic reviews have covered key areas such as distributed energy resources (Evans et al. 2012), sustainable development (Dincer and Acar 2015), energy policy (Jenkins et al. 2016), solar and wind power (Khare et al. 2016), energy efficiency (Olatomiwa et al. 2016), flexibility requirements (Kondziella and Bruckner 2016), and multi-energy systems (Mancarella 2014) with individual reviews over the last decade, there are far fewer review type contributions covering the broad field of energy systems analysis in its entirety. When it comes to the research methods, previous investigations have focused on optimisation techniques (Baños et al. 2011), multi-criteria decision-making (Kumar et al. 2017), agent-based modelling (Scheller et al. 2019), stochastic planning (Sharma et al. 2012) and life cycle assessments (Martín-Gamboa et al. 2017). These publications

---

[1] Scopus search query for reviews, which is the same as the search query for articles of this study: TITLE-ABS-KEY ( ( "energy system*" ) AND ( simul* OR model* OR "plan" OR "plans" OR "planning" OR optimi* OR analy* OR assess* OR evaluat* ) ) AND ( LIMIT-TO ( SUBJAREA , "ENGI" ) OR LIMIT-TO ( SUBJAREA , "ENER" ) OR LIMIT-TO ( SUBJAREA , "ENVI" ) ) AND ( LIMIT-TO ( DOCTYPE , "ar" ) ) AND PUBYEAR > 2010 AND ( LIMIT-TO ( LANGUAGE , "English" ) )

also cover different subject areas such as Energy, Engineering, Environmental Sciences, Social Sciences, Chemical Engineering, Material Science, Mathematics, and Business Management.

Rather than focus on a specific method or area of application, this paper takes a broader perspective. Precisely because of the breadth of the field of ESA and the amount of research activity, this paper deals with mapping competencies, collaborations and output across the whole field over the last two decades. The employed methods are a combination of literature review and bibliometrics, whereby the precise methodology and search terms employed can be found in Section 2.

Based on a review of the literature as discussed above and throughout the article, with a focus on the field of energy systems analysis, the central research questions of this paper are:

- What key topics and dynamic trends can be identified, and which are emerging or could be emerging for the coming years?
- Which authors/institutions/countries collaborate most?
- Which authors/institutions/countries are most productive over time, and what indicators can be employed to measure this?

The paper is structured as follows: Section 2 explains the employed methodology before Section 3 presents the results. Section 4 discusses the implications of the results, while a summary and conclusions are presented in Section 5.

## 2. Methodology

An overview of the methods and results of this paper can be seen in Figure 1, whereby the methods comprise of different statistical techniques to investigate the structure, characteristics and patterns of the underlying science (Weinand 2020a). The bibliographic database *Scopus (Elsevier 2020)* was employed to analyse the literature on ESA (Section 2.1). Alongside the web interface *biblioshiny* of the R-tool *bibliometrix* (Aria and Cuccurullo 2017) (Section 2.2), the statistical indicators h-, m- and g-indices (Section 2.3), a method for measuring trends (Section 2.4), an approach for analysing a countries' impact (Section 2.5) and an algorithm for the investigation of keywords (Section 2.6) are explained in the following subsections.

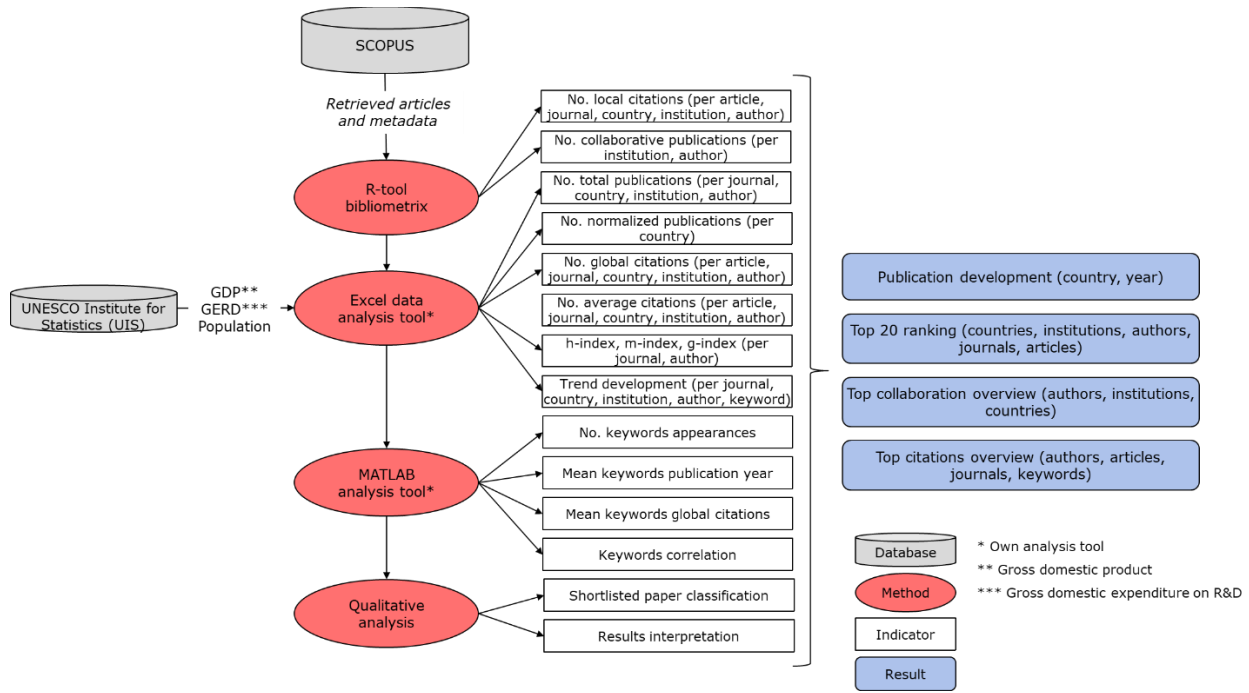

*Figure 1. Logical tree representation of the methods used in the paper. In addition to existing tools, own tools have been applied for the analysis of the retrieved articles.*

### 2.1. Bibliometrics search terms

The corpus of documents used in the bibliometric analysis was the result of a thorough filtering process. From the field of *energy*, the search was restricted to documents related to the analysis and modelling of energy systems in the last 20 years, as shown by the *initial search* in Table 1. The search resulted in 30,391 documents of which 14,215 were articles in English. Furthermore, to exclude results that were considered unsuitable for the ESA area, the search was narrowed to the following research areas: Engineering, Environment and Energy. The final search terms are presented in the *adjusted search* row in Table 1.

*Table 1: Search queries and the resulting number of articles in the literature database Scopus.*

| Search name | Search query | Date | Number of documents |
|---|---|---|---|
| "Energy" search | TITLE-ABS-KEY (energy) AND PUBYEAR > 1999 AND PUBYEAR < 2020 AND (LIMIT-TO (SUBAREA "ENGI") OR LIMIT-TO (SUBAREA "ENER") OR LIMIT-TO (SUBAREA "ENVI")) AND (LIMIT-TO (DOCTYPE "ar")) AND (LIMIT-TO (LANGUAGE "English")) | 25.05.2020 | 762,418 |
| Initial search | TITLE-ABS-KEY (("energy system*") AND (simul* OR model* OR "plan" OR "plans" OR "planning" OR optimi* OR analy* OR assess* OR evaluat*)) AND PUBYEAR > 1999 AND PUBYEAR < 2020 | 25.05.2020 | 30,391* |
| Adjusted search | TITLE-ABS-KEY (("energy system*") AND (simul* OR model* OR "plan" OR "plans" OR "planning" OR optimi* OR analy* OR assess* OR evaluat*)) AND PUBYEAR > 1999 AND PUBYEAR < 2020 AND (LIMIT-TO (SUBAREA "ENGI") OR LIMIT-TO (SUBAREA "ENER") OR LIMIT-TO (SUBAREA "ENVI")) AND (LIMIT-TO (DOCTYPE "ar")) AND (LIMIT-TO (LANGUAGE "English")) | 25.05.2020 | 12,182* |

\* Out of initial 30,391 documents, 14,450 were articles and 28,845 were publications in English. After limiting the search to articles in English, the filtered number of publications was 14,215. After limiting to the subject areas (to exclude results that we assessed that were not suitable for our analysis) the 12,182 publications were left.

## 2.2. R-tool bibliometrix

Once the data has been retrieved from the Scopus database, the R-Tool bibliometrix was applied to examine the corpus of literature. Bibliometrix is an open-source tool for conducting comprehensive scientific mapping analyses. This tool is implemented in R, and therefore the package is flexible and facilitates integration with other statistical or graphical packages (Aria and Cuccurullo 2017). One example for which bibliometrix was used is the number of collaborative publications of countries (Table 3)

## 2.3. Measures of influence: h-index, m-index and g-index

Several measures were employed in order to assess researchers', institutions' and countries' output productivity. The h-index quantifies the cumulative impact and relevance of an individual's scientific output (Hirsch 2005), whereby an individual here denotes researchers, institutions and countries. It reflects the number of $h$ papers of an individual that have been cited at least $h$ times. For example, if 50 publications of an author have at least 50 citations, then that author's h-index is 50.

In Hirsch (2005), the m-index was also introduced, which reflects the period since the first publication of an individual by dividing the h-index by the number of years of scientific activity. For example, if an individual has an h-index of ten after twenty years of scientific activity, then his m-index would be 0.5. According to Hirsch (2005), an individual with an m-index of $m = 1$ is a "successful scientist", with $m = 2$ an "outstanding scientist" and with $m = 3$ a "truly unique individual".

The g-index was suggested by Egghe (2006), as an alternative to the h-index. It represents the unique largest number of the top $g$ most cited articles, which together received at least $g^2$ citations. Therefore, highly cited articles are weighted stronger than with the h-index. For example, five publications of which one is cited 21 times and the others only once each (25 citations in sum) result in a g-index of five and an h-index of only one.

In this study, the indices presented refer only to the analysed literature body (based on the search terms outlined above). This is because publications on ESA often only constitute a subset of an individual's total publications and cannot be used to measure the general scientific influence of an individual.

## 2.4. Measuring trends

In order to measure the trends in publication numbers of authors, institutions or countries as well as on specific topics, the development of the last five years (2015-2019) was analysed. Thereby the slope in publications per year was used. For example, if an author published one article in 2015, two in 2016, three in 2017, four in 2018 and five in 2019 the slope is 1.0. After determining the publication development (slope) for all individuals, five ranges of trends were introduced in order to obtain two ranges of positive and negative trends and one range for a steady publication number (i.e. slope = 0.0). Table 2 shows an example for the case with the maximum slope of one of the individuals of 70.4 and a minimum slope of -10.6.

*Table 2: Range intervals and corresponding arrow signs for indicating the trend in publication numbers for example with a maximum slope in publication numbers of 70.4 and a minimum slope of -10.6.*

| Range interval | Arrow for indicating the trend |
|---|---|
| [-10.6; -5.3) | ↓ |
| [-5.3; 0.0) | ↘ |
| [0.0] | → |
| (0.0; 35.2] | ↗ |
| (35.2; 70.4] | ↑ |

## 2.5. Analysis of the countries' impact

In order to compare the research impact of different countries, several indexes can be used. To narrow the scope of the analysis, the number of analysed countries was reduced to 20 based on their total number of publications within the timeframe 2000-2020. However, this approach is biased towards large countries. Furthermore, Man et al. (2004) and Meo et al. (2013) identified a positive correlation between the Gross Domestic Expenditure on R&D (GERD) and the country's total number of publications. Therefore, in order to level off the differences between the countries, their publication output was normalised based on three factors: the GERD, the Gross Domestic Product (GDP) and the population (cf. Section 3.1).

## 2.6. Keyword analysis

Further analyses are performed with a MATLAB algorithm, which was based on Weinand et al. (2020b). In this algorithm, the similarity of strings is determined with the Levenshtein distance (Levenshtein 1966). Thereby, the number of occurrences of author keywords could be measured by matching equivalent strings under one keyword. This ensured that slightly different keywords, such as the plural of a word or the use of a hyphen, are recognised as the same keyword (e.g. smart grid and smart-grids). In addition to the number of occurrences of the keywords, the algorithm also calculated the mean publication year and the mean citations of all articles containing the respective keyword. Furthermore, the algorithm examined the keywords for their simultaneous occurrence in the same articles. Further brief analyses were performed with additional simple methods and data, which is explained in Section 3 next to the relevant results.

# 3. Results

The results are presented in this section as follows. First, the focus is on the national publication output and productivity in Section 3.1, followed by institutional output and collaboration in Section 3.2. In Section 3.3, attention is then focussed on the productivity of individual authors and their key collaborations. Section 3.4 subsequently examines different journals and individual articles, before Section 3.5 explores the keywords and central themes within the field of ESA.

### 3.1. National publication output and productivity

Figure 2 presents the development of publications within the field of ESA during the last two decades. The number of total publications showed quasi exponential growth during the period, and the sharp increase between 2016 and 2019 underlines the increasing importance of ESA. Compared to the general energy-related literature, there was a greater increase in publications in the ESA field, especially in the last five years. The growth of ESY publications from 2015 to 2019 was 242% (940 to 2276), while the general energy-related publications saw a growth of 161% (57,385 to 92,632). Furthermore, the analysis of search queries used in Table 1 shows that the *Energy system analysis* articles increased their share in the articles published in the general field of *Energy* between the years 2000 and 2019. In the year 2000, the share of *Energy system analysis* publications in *total energy publications* was 0.6%, while in the year 2019, its share increased to 2.46%.

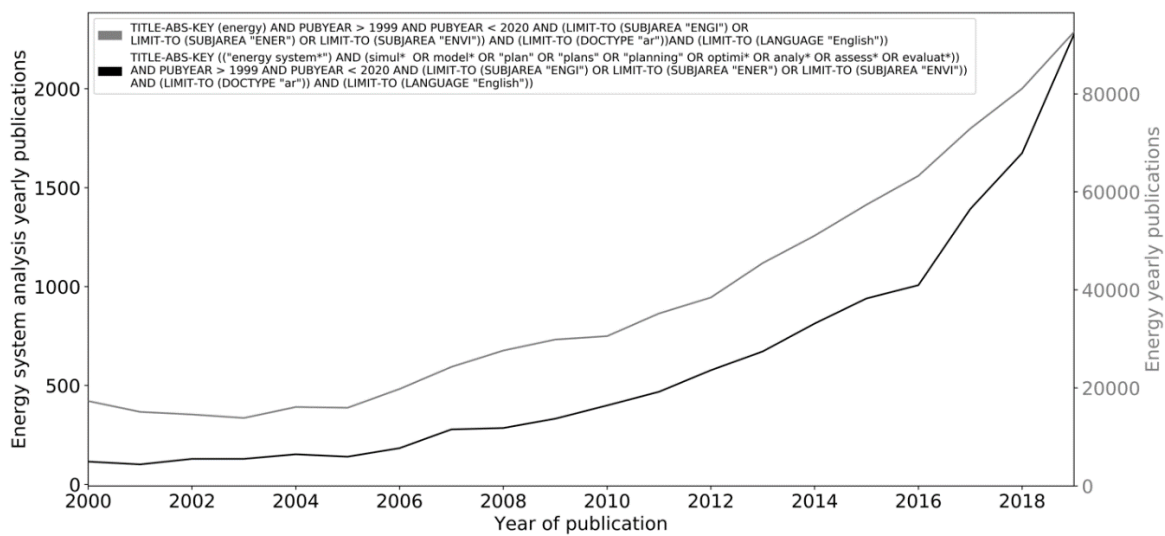

*Figure 2: Number of publications per year for the Energy system analysis field ("adjusted search" from Table 1) and the Energy field ("Energy search" from Table 1) over the period 2000-2019.*

Figure 3 gives an overview of the productivity of the top 20 countries in terms of the total number of publications. Although the United States of America (USA) was the country with the largest number of publications, its relative production in relation to the GDP and the total population was rather low. A similar pattern of very low relative indicators can also be seen for China, India and Japan. The United Kingdom and Iran had a very high number of publications relative to Gross domestic expenditure on R&D (GERD). Relative to the GDP and the total population, Denmark, Sweden, Switzerland, Austria, the Netherlands and Finland significantly outperformed the other countries on the list.

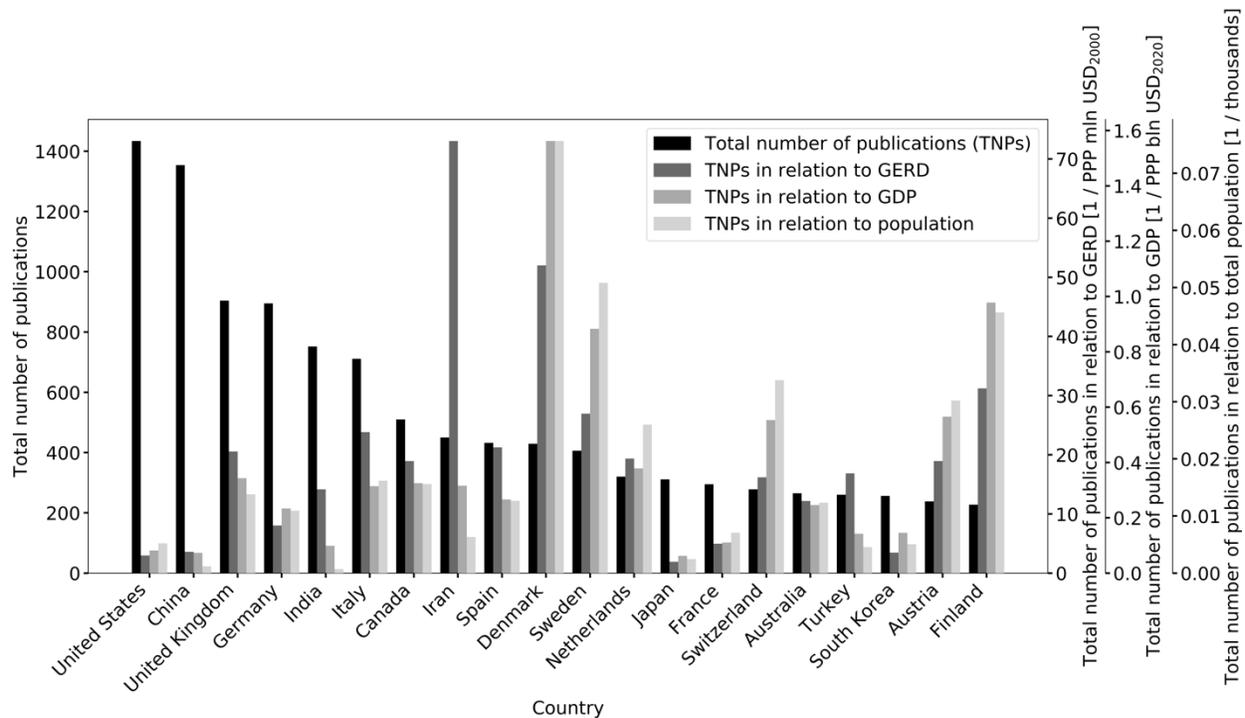

*Figure 3: The top 20 most productive countries in terms of the total number of publications on energy system analysis in the period 2010-2019, that is a subset of the whole analysed period from 2000 to 2019 as presented in Table 4. On the right side, the secondary y-axes show the normalised total number of publications based on the GERD[2], GDP[3] and the total population for each country. For each normalisation factor, the mean value in the range of 2010-2019 is used.*

Table 3 presents the top 20 most productive countries in the field of ESA, measured by the total number of publications. The top 5 countries (USA, China, United Kingdom, Germany, Italy) covered about 50% of the total publications. The USA tops the list both in the total publications indicator, as well as in the country of the first author's affiliation and in terms of all h, g and m indices.

In terms of relevance in the ESA field, the most productive countries in terms of Total Number of Publications (TNPs) will still be the most relevant ones, independently of their size. In line with the yearly growing TNPs, as shown in Figure 2, there is also a positive trend regarding the number of publications in each listed country (Table 3). While nearly all countries show a similar average increase in publications from 2015 to 2019, China represents a positive outlier. Accordingly, it is expected that China will replace the USA at the top in the next few years concerning the TNPs.

The most productive countries in the field in terms of the TNPs are very populous countries, including the USA, China and India. However, Sweden and Denmark managed to enter the top 10 most productive countries list, showing that the productivity of different countries in this field is not completely correlated.

---

China and the United Kingdom were the most productive countries in terms of the number of publications per expenditure on research and development in general.

Looking into Figure 3 and Table 3, China, the UK and Iran had a significantly higher TPGD indicator, meaning that their funding was used more efficiently or that they invested a larger share of the total R&D spending to the energy system analysis field.

Throughout this manuscript, a simple correlation analysis was done for many instances. The Pearson correlation coefficient r was used in all instances. The number of observations was always 20, and the correlation coefficient was deemed positive if the R-value of the correlation coefficient was greater than or equal to 0.38, with a statistically significant value that was determined as a p-value of less than 0.1.

Simple correlation analyses between the h-, g- and m-indices presented in Table 3 and various normalised indicators presented in Figure 3 revealed that the GERD indicator is a reliable and relevant metric when evaluating the impact of the research, as well as general research productivity, contrary to the general productivity of society expressed via GDP. While the normalised TNPs based on the GERD was correlated with all of the h, g and m indices (correlation coefficient from 0.56 to 0.66), the impact of publications expressed via h-, g- and m-indices was uncorrelated (correlation coefficient less than 0.38) with both GDP and the total population. Furthermore, the TNPs was correlated to the GERD indicator but was uncorrelated to both the GDP and the Total population. The presented findings confirmed the findings from Meo et al. (2013), in this paper, more specifically for the ESY field. Meo et al. (2013) showed that the GERD is correlated with the total number of published documents which is similar to TNPs, citations per documents and h-index, while they did not find a strong correlation between GDP per capita and research outcomes, while their study was carried out for a wide range of social and science subjects. Further correlation analyses in relation to Table 3 showed that there was no correlation between the share of collaborative publications and the TNPs nor any of the indices. The latter points to the conclusion that so far in the field of ESA, transnational collaborative approaches did not generate more impactful papers on average. On the other hand, there was a strong correlation between the TNPs and all three measures of influence, with correlation coefficients between 0.7 and 0.82. The latter points to the conclusion that, on average, the most productive countries also generate the most impactful papers in the field, showing that the quantity and the impact of papers are highly correlated in this field.

Table 3: The top 20 of the most productive countries in terms of publications on energy system analysis. The numbers for "total publications" do not have to sum up to 12,182 or 100%, since more than one country could have contributed to a single publication. The percentages for "total publications" and "corresponding author's country" refer to the total number of 12,182 articles, while for "single publication" and "collaborative publication" they refer to the number of publications of "corresponding author's country". *In the calculation of the "average number of authors per publication" it is excluded the document Aad et al. (2013). It is a joint publication between 3060 authors. A large number of authors would skew the average, making the comparison between countries less meaningful.

| Country | Total number of publications (TNPs) | | Trend | h-ind. | g-ind. | m-ind. | Corresponding author's country | | Single country publication | | Collaborative publications | | Average number of authors per publication* |
|---|---|---|---|---|---|---|---|---|---|---|---|---|---|
| | No. | % | | | | | No. | % | No. | % | No. | % | |
| USA | 1,803 | 15 | ↗ | 100 | 155 | 5.00 | 927 | 7.6 | 761 | 82 | 166 | 18 | 4.78 |
| China | 1,446 | 12 | ↑ | 68 | 97 | 3.40 | 559 | 4.6 | 398 | 71 | 161 | 29 | 4.81 |
| United Kingdom | 1,031 | 8 | ↗ | 79 | 120 | 3.95 | 462 | 3.8 | 353 | 76 | 109 | 24 | 4.71 |
| Germany | 1,019 | 8 | ↗ | 71 | 110 | 3.55 | 594 | 4.9 | 469 | 79 | 125 | 21 | 5.15 |
| Italy | 818 | 7 | ↗ | 63 | 91 | 3.15 | 505 | 4.1 | 400 | 79 | 105 | 21 | 4.66 |
| India | 805 | 7 | ↗ | 45 | 67 | 2.25 | 375 | 3.1 | 337 | 90 | 38 | 10 | 3.18 |
| Canada | 596 | 5 | ↗ | 65 | 100 | 3.25 | 320 | 2.6 | 228 | 71 | 92 | 29 | 3.88 |
| Sweden | 531 | 4 | ↗ | 59 | 98 | 2.95 | 277 | 2.3 | 219 | 79 | 58 | 21 | 4.21 |
| Spain | 515 | 4 | ↗ | 57 | 102 | 2.85 | 290 | 2.4 | 201 | 69 | 89 | 31 | 4.90 |
| Denmark | 494 | 4 | ↗ | 75 | 126 | 3.75 | 289 | 2.4 | 208 | 72 | 81 | 28 | 4.58 |
| Japan | 493 | 4 | ↗ | 46 | 70 | 2.30 | 357 | 2.9 | 299 | 84 | 58 | 16 | 4.64 |
| Iran | 463 | 4 | ↗ | 51 | 74 | 3.19 | 164 | 1.3 | 135 | 82 | 29 | 18 | 3.55 |
| Netherlands | 381 | 3 | ↗ | 52 | 89 | 2.60 | 206 | 1.7 | 124 | 60 | 82 | 40 | 5.40 |
| France | 354 | 3 | ↗ | 45 | 79 | 2.25 | 140 | 1.1 | 88 | 63 | 52 | 37 | 5.08 |
| Turkey | 336 | 3 | ↗ | 46 | 79 | 2.56 | 170 | 1.4 | 143 | 84 | 27 | 16 | 2.86 |
| Switzerland | 331 | 3 | ↗ | 52 | 76 | 2.60 | 167 | 1.4 | 103 | 62 | 64 | 38 | 4.56 |
| Australia | 294 | 2 | ↗ | 43 | 75 | 2.15 | 100 | 0.8 | 81 | 81 | 19 | 19 | 4.08 |
| Austria | 287 | 2 | ↗ | 52 | 96 | 2.60 | 139 | 1.1 | 79 | 57 | 60 | 43 | 5.80 |
| South Korea | 270 | 2 | ↗ | 34 | 57 | 1.89 | 226 | 1.9 | 172 | 76 | 54 | 24 | 4.28 |
| Finland | 258 | 2 | ↗ | 34 | 56 | 1.70 | 164 | 1.3 | 134 | 82 | 30 | 18 | 4.39 |

## 3.2. Institutional output and collaboration

The ordered list of institutions per total publications can be seen in Table 4. The first two positions are both held by Danish universities, which were affiliated with 3.6% of the publication in this research field and published 88% of the total publications in the country. The outstanding position of both institutions is also underlined by a consistently high h-, g-, and m-index. The two Swedish universities Royal Institute of Technology KTH and Chalmers University of Technology were affiliated with 50% of the total publications of Sweden in the field, while the UK's universities Imperial College London and the University College London were responsible for 29% of the UK publications in the field. The Aalto University, the International Institute for Applied Systems Analysis and the ETH Zurich took similar outstanding roles for their respective countries (Finland, Austria and Switzerland respectively) by accounting for at least 37% of the total publications of their countries in this field. Finally, four of the Chinese institutions that made it onto the list (Tsinghua University, North China Electric Power University, Chinese Academy of Science and the Ministry of Education China) produced 38% of the Chinese publications in the field. This analysis shows that only a few most productive institutes are responsible for the bulk of the publications within the most productive countries, meaning that the development of the field is highly concentrated within a few groups and institutions.

Although the USA was the most productive country in terms of total publications, it does not have an institution on the list of top 20 institutions. Interestingly, the first USA institute is NASA with 82 publications. On the other hand, four institutes are based in China. This shows that the research groups and their financing is more distributed in the USA than elsewhere. At the same time, the influence indices of the Chinese institutions were closer to the lower midfield.

*Table 4: The top 20 of the most productive institutions in terms of publications on energy system analysis. The percentage values refer to a total of 12,182 publications. *In the calculation of the "average number of authors per publication" it is excluded the document Aad et al. (2013). It is a joint publication between 3060 authors. Such a large number of authors would skew the average, making the comparison between countries less meaningful.*

| Institution | Country | Total publications | | Trend | h-index | g-index | m-index | Average number of authors per publication* |
|---|---|---|---|---|---|---|---|---|
| | | No. | % | | | | | No. |
| Aalborg University | Denmark | 242 | 2.0 | ↑ | 63 | 113 | 3.15 | 3.51 |
| Denmark Technical University | Denmark | 193 | 1.6 | ↑ | 43 | 72 | 2.15 | 4.00 |
| Tsinghua University | China | 169 | 1.4 | ↑ | 34 | 58 | 1.79 | 4.24 |
| Imperial College London | United Kingdom | 161 | 1.3 | ↑ | 39 | 64 | 2.79 | 4.94 |
| North China Electric Power University | China | 149 | 1.2 | ↑ | 24 | 37 | 1.20 | 4.42 |
| Ontario Tech University | Canada | 137 | 1.1 | ↗ | 40 | 66 | 2.35 | 2.95 |
| Chinese Academy of Science | China | 137 | 1.1 | ↗ | 28 | 48 | 1.75 | 4.00 |
| University College London | United Kingdom | 135 | 1.1 | ↗ | 31 | 51 | 2.58 | 4.89 |
| The Royal Institute of Technology KTH | Sweden | 134 | 1.1 | ↗ | 28 | 51 | 1.56 | 5.18 |
| Chalmers University of Technology | Sweden | 132 | 1.1 | ↗ | 33 | 64 | 1.65 | 3.90 |
| ETH Zürich | Switzerland | 125 | 1.0 | ↗ | 35 | 58 | 1.94 | 4.00 |
| International Institute for Applied Systems Analysis | Austria | 110 | 0.9 | → | 39 | 60 | 1.95 | 6.30 |
| Aalto University | Finland | 108 | 0.9 | ↗ | 23 | 37 | 1.44 | 4.20 |
| RWTH Aachen | Germany | 102 | 0.8 | ↑ | 28 | 44 | 1.40 | 5.34 |
| Ministry of Education China | China | 99 | 0.8 | ↑ | 20 | 33 | 1.43 | 4.00 |
| National Technical University of Athens | Greece | 99 | 0.8 | ↗ | 28 | 46 | 1.40 | 3.77 |
| Norwegian University of Science and Technology | Norway | 94 | 0.8 | ↗ | 28 | 49 | 1.47 | 4.65 |
| University of Tehran | Iran | 92 | 0.8 | ↑ | 28 | 42 | 2.33 | 3.56 |
| Polytechnic University of Turin | Italy | 91 | 0.7 | ↗ | 25 | 43 | 1.25 | 4.00 |
| University of Tokyo | Japan | 87 | 0.7 | ↗ | 23 | 37 | 1.15 | 3.45 |

The collaboration network of institutions, presented in Figure 4, revealed that most of the influential institutions in the ESA field collaborate overwhelmingly within the country. This could be because of the energy policy frameworks that can significantly differ from country to country, funding regimes that favour domestic institutions, educations system specifics, and other reasons. Institutions from the United Kingdom, Netherlands, Austria, Sweden, and Canada seem a bit more open to collaboration with other countries. Very active collaboration exists between the RWTH Aachen and E.On. Furthermore, the Ministry of Education China and the Tianjin University, the University of Regina, the North China Electric Power University, and RWTH Aachen and the Forschungszentrum Julich FZJ had a high number of publications in common.

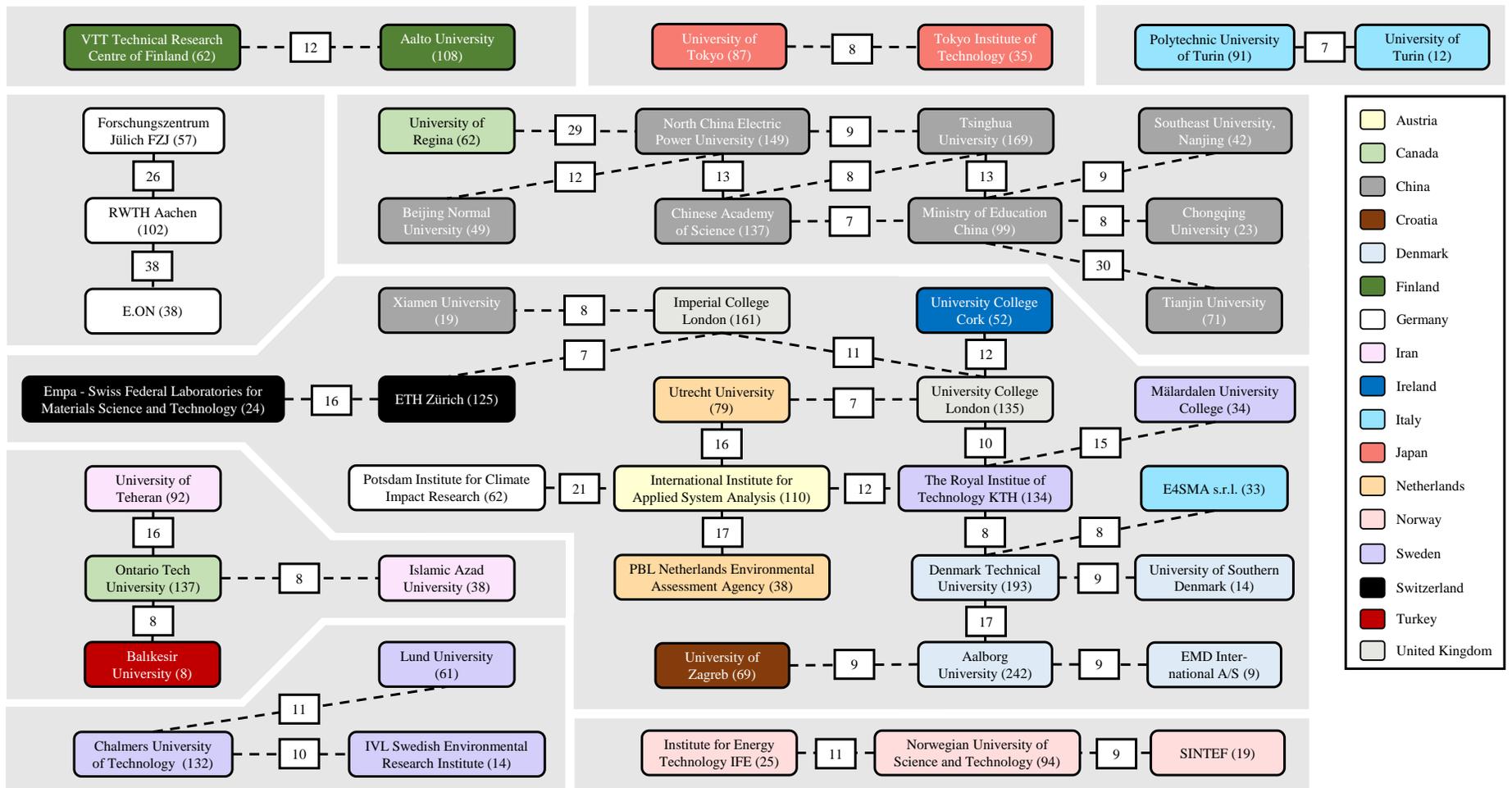

*Figure 4: Collaboration network of the top 20 most productive institutions, together with their top three most productive collaborations. Collaborations with less than five collaborative publications are not shown. The number in brackets after the institute describes the number of total publications within the ESA corpus analysed in this paper. The number in the rectangle on the connection between two institutes indicates the joint number of publications.*

A look at the collaboration networks of the most relevant authors (Figure 5) reveals several points. First, the core of their collaborations were colleagues from the same groups or institutions. For example, the connection between the top authors Lund H and Mathiesen BV is emphasised with 27 joint publications. Furthermore, Lund H published 45 publications with Mathiesen BV, Østergaard PA and Möller B, which belong to the same institution. The same goes for the cooperation of the top authors Dincer I and Rosen MA as well as Li YP and Huang GH with each of the pair of authors having 21 joint publications. Second, the most productive authors also had one to two dominant collaboration networks in other institutions. For the case of Dincer I and Rosen MA, those were the colleagues from the University of Tehran (Ahmadi P and Maleki A) and Ege University (Hepbasli A), for the case of Lund, that was the University of Zagreb (Duić N and Krajačić G) and the Technical University of Denmark (Münster M and Karlsson K), while for the case of Huang GH, it was the University of Regina (Lin QG). It is interesting to note that the current affiliation of Huang GH is the University of Regina and Lin QG's current affiliation is the North China Electric Power University, meaning that those two authors exchanged positions between the same universities during their careers, which might also be one reason for the strong collaboration. Third, the list of collaboration networks of the authors presented in Figure 5 also confirmed that the most dominant collaboration networks were within institutions coming from the same country.

The list of the most productive authors showed significant differences according to the different indices and the total number of relevant publications (Table 5). It should be noted here that the list was created based solely on the top 20 authors according to the total number of relevant publications and not according to the other indices. According to the total publications, the top 3 authors were Dincer I (University of Ontario Institute of Technology), Lund H (Aalborg University), and Rosen MA (University of Ontario Institute of Technology). While the University of Ontario had two authors in the top 3, the Aalborg University exhibited three authors in the top 6. Lund H topped the ranking in terms of the three indices, being the first based on h- and g-indices, and the third based on the m-index. Furthermore, the performance over the last years of Breyer C (Lappeenranta University of Technology) has been promising since he has been cited most frequently in the last years (m-index of 3.00), which is significantly higher than from the other authors in the list, closely followed by Mathiesen BV (Aalborg University, m-index of 2.58). The correlation analysis of the authors' indices revealed that the h- and g-indices were mutually highly correlated (r-value of 0.96), while h- and m- and g- and m- indices were only weakly correlated (r values of 0.48 and 0.38 respectively). Moreover, the total number of publications of the most productive authors was highly correlated with h- and g-indices (r values of 0.88 and 0.93, respectively), while it was uncorrelated with the m-index (0.35). This shows that the most productive authors also had the largest impact in terms of h- and g-indices. Finally, the average number of authors per publication was negatively correlated with h-, g- and m-indices (-0.54, -0.48 and -0.38, respectively). The latter shows that the most productive authors tended to collaborate in smaller groups. It is interesting to note that the top three authors (Dincer I., Lund H. and Rosen MA) all tended to collaborate in smaller groups with an average number of authors of 2.7, 3.3 and 3.3 respectively.

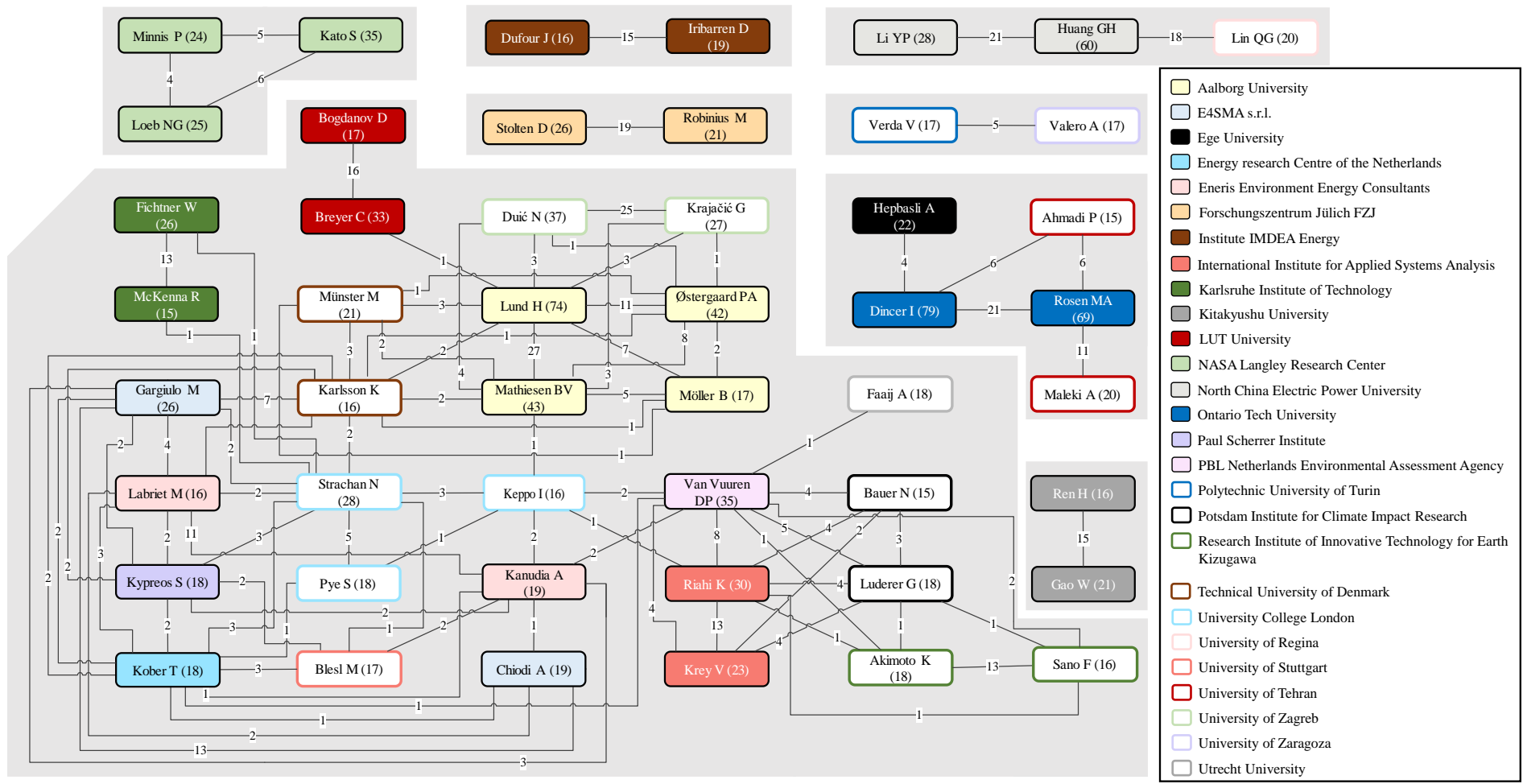

*Figure 5: Collaborations network of authors with at least 15 publications and at least one collaboration among these authors with at least three collaborative publications. In parentheses behind the authors is the total number of publications on ESA, and the numbers on the edges refer to the number of collaborations. The colours show the affiliation of the authors for the majority of their publications.*

Table 5: The top 20 of the most productive authors in terms of publications on energy system analysis.

| Author | Total publications | Trend | h-index | g-index | m-index | The average number of authors per publication | Histogram |
|--------|-------------------|-------|---------|---------|---------|----------------------------------------------|-----------|
| *Dincer I* | 79 | ↘ | 32 | 55 | 1.68 | 2.70 | |
| *Lund H* | 74 | ↗ | 43 | 74 | 2.15 | 3.30 | |
| *Rosen MA* | 69 | ↗ | 31 | 57 | 1.72 | 3.30 | |
| *Huang GH* | 60 | ↘ | 24 | 39 | 1.60 | 4.17 | |
| *Mathiesen BV* | 43 | ↗ | 31 | 43 | 2.58 | 4.37 | |
| *Østergaard PA* | 42 | ↗ | 20 | 39 | 1.00 | 2.85 | |
| *Duić N* | 37 | ↓ | 24 | 37 | 1.41 | 4.76 | |
| *Breyer C* | 33 | ↑ | 18 | 29 | 3.00 | 4.03 | |
| *Maréchal F* | 32 | ↘ | 16 | 27 | 1.15 | 3.53 | |
| *Müller D* | 30 | ↗ | 15 | 22 | 1.88 | 5.03 | |
| *Riahi K* | 30 | ↗ | 19 | 30 | 1.19 | 9.13 | |
| *Li YP* | 28 | ↗ | 17 | 27 | 1.71 | 3.89 | |
| *Strachan N* | 28 | ↓ | 17 | 28 | 1.42 | 5.11 | |
| *Howells M* | 27 | ↘ | 12 | 26 | 1.20 | 6.85 | |
| *Krajačić G* | 27 | ↘ | 19 | 27 | 1.46 | 5.07 | |
| *Mancarella P* | 27 | ↗ | 18 | 27 | 1.29 | 3.04 | |
| *Fichtner W* | 26 | ↗ | 15 | 23 | 0.79 | 4.15 | |
| *Gargiulo M* | 26 | ↗ | 13 | 24 | 1.44 | 7.04 | |
| *Stolten D* | 26 | ↑ | 10 | 24 | 1.25 | 6.46 | |
| *Loeb NG* | 25 | ↘ | 14 | 25 | 0.78 | 5.84 | |

### 3.3. Journals and articles

The *Energy* journal was the most productive one in terms of the total number of relevant publications with 1,017 publications, while its h-, g- and m- indices were 77, 117 and 3.67, respectively (Table 6). However, according to all the other indices, the journals *Applied Energy* (h- and m-index of 86 and 4.1, respectively) and *Energy Policy* (g-index of 137) outperformed it. The relatively young journal *Energy Research and Social Science* (established in 2014) also showed an impressive m-index of 4.0, demonstrating that the social science topics within the field of ESA have been gaining more importance. Moreover, based on the m-index, it is expected that the *Applied Energy*, and *Energy Research and Social Science* journals will have the most impactful publications in the future. Although the journal *Energies* reached a high place on the list according to the total relevant publications with 460 publications, its number of average article citations was very low (7 per article). The situation was even worse for the journal *Renewable Energy and Power Quality Journal* (with two citations on average per article). The first six journals from Table 6 published 30.4% of the total publications in the field. However, all the top 20 journals published only 47.1% of the publication in the field. The latter shows that the publications are very spread among the many different journals, making it increasingly hard to follow the development in the field, potentially reducing and slowing down the spread of research results.

A correlation analysis showed that the measures of influence, expressed as h-, g- and m-indices, were highly correlated with the total number of publications in different journals (r values of 0.83, 0.79 and 0.871 respectively), similarly to the findings relevant for different countries and their productivity. However, average article citations were very weekly correlated with the total number of publications, with the correlation coefficient of 0.44. As h-, g- and m-indices are based on the absolute numbers, their values will naturally be higher for journals that have more publications, as the indices do not include a penalty for low-quality articles. A very low correlation coefficient between the total number of citations and the number of articles published in specific journals reveal that the quality of articles in the field of ESA is not very concentrated in the small number of journals.

Table 6: The top 20 of the most productive scientific journals in terms of publications on energy system analysis. The percentage values refer to a total of 12,182 publications.

| Journal | Total publications (TP) | | Trend | Average article citations | h-index | g-index | m-index | Average number of authors per publication |
|---|---|---|---|---|---|---|---|---|
| | No. | % | | | | | | No. |
| *Energy* | 1,017 | 8.3 | ↑ | 30 | 77 | 117 | 3.67 | 3.61 |
| *Applied Energy* | 846 | 6.9 | ↑ | 37 | 86 | 124 | 4.10 | 3.87 |
| *Energy Policy* | 603 | 4.9 | ↗ | 47 | 77 | 137 | 3.67 | 3.33 |
| *Energies* | 460 | 3.8 | ↑ | 7 | 23 | 31 | 2.30 | 4.17 |
| *Renewable Energy* | 421 | 3.5 | ↗ | 42 | 73 | 109 | 3.48 | 3.31 |
| *Energy Conversion and Management* | 367 | 3.0 | ↗ | 33 | 57 | 86 | 2.71 | 3.48 |
| *International Journal of Hydrogen Energy* | 311 | 2.6 | ↗ | 31 | 47 | 82 | 2.35 | 3.74 |
| *Journal of Cleaner Production* | 237 | 1.9 | ↗ | 22 | 36 | 58 | 2.00 | 3.84 |
| *Energy and Buildings* | 208 | 1.7 | ↓ | 29 | 41 | 63 | 1.95 | 3.68 |
| *Applied Thermal Engineering* | 170 | 1.4 | ↘ | 29 | 38 | 60 | 1.81 | 3.55 |
| *Solar Energy* | 170 | 1.4 | ↗ | 40 | 47 | 77 | 2.24 | 3.45 |
| *Sustainability* | 155 | 1.3 | ↗ | 7 | 16 | 23 | 2.00 | 2.47 |
| *Energy Research and Social Science* | 117 | 1.0 | ↗ | 22 | 28 | 46 | 4.00 | 2.86 |
| *International Journal of Energy Research* | 116 | 1.0 | ↗ | 19 | 28 | 40 | 1.33 | 3.11 |
| *International Journal of Renewable Energy Research* | 107 | 0.9 | ↗ | 8 | 15 | 21 | 1.50 | 3.00 |
| *Energy Strategy Reviews* | 97 | 0.8 | ↗ | 12 | 20 | 29 | 2.22 | 4.88 |
| *International Journal of Electrical Power and Energy Systems* | 93 | 0.8 | ↗ | 27 | 31 | 45 | 1.63 | 3.42 |
| *Renewable Energy and Power Quality Journal* | 80 | 0.7 | ↘ | 2 | 6 | 7 | 0.33 | 3.44 |
| *Energy Economics* | 77 | 0.6 | ↘ | 35 | 32 | 49 | 1.52 | 3.74 |
| *Sustainable Cities and Society* | 76 | 0.6 | ↗ | 15 | 21 | 28 | 2.10 | 3.46 |

The list of the most frequently cited papers is dominated by the papers that are ten or more years old (Table 7) The most recently published paper on the list is Connolly et al. (2014). Palensky and Dietrich (2011) was the most cited paper from the list and has been also cited the most on *per year* basis.

In terms of the top 20 most cited articles, five articles were published in the journal Energy Policy (Connolly et al. 2014; Unruh 2000; Jacobson and Delucchi 2011; Lund and Kempton 2008; Jacobsson and Lauber 2006; Jacobsson and Johnson 2000), three articles were published in the journal Energy (Lund 2007; Lund and Mathiesen 2009; Lund et al. 2010), and one publication appeared in Applied Energy (Kalogirou 2000) and Solar Energy (Yang et al. 2007) respectively. Lund H wrote four out of 20 top publications as a first author (Lund 2007; Lund and Mathiesen 2009; Lund et al. 2010; Lund 2005). Mathiesen BV was involved in two collaborations with Lund H (Lund and Mathiesen 2009; Mathiesen et al. 2011). In this context, from the top institutions, Aalborg University represents four (Lund 2007; Lund and Mathiesen 2009; Lund et al. 2010; Holm-Nielsen et al. 2009), the Chalmers University of Technology two (Jacobsson and Lauber 2006; Jacobsson and Johnson 2000), and Denmark Technical University represents one (Holm-Nielsen et al. 2009) of the most cited publications in this field. At the same time, six of the most referenced articles came from the USA (Jacobson and Delucchi 2011; Lund and Kempton 2008; Crawley et al. 2008; Zhang et al. 2004; Thomson et al. 2011; Wang and Nehrir 2008), followed by Denmark with five (Lund and Kempton 2008; Lund 2007; Lund and Mathiesen 2009; Lund et al. 2010; Holm-Nielsen et al. 2009). These paragraphs showed that the impact of specific authors, institutions and countries in the ESA research field depends on more factors than productivity as analysed above (Figure 3 and Table 3), which illustrates the importance of employing multiple indicators.

*Table 7: Most frequently cited articles among the scientific contributions on ESA. The articles are sorted by the number of global citations. At the end of the table, the six articles from the top 10 of the most locally cited articles are added, which were not already contained in the top 20 of the most globally cited articles.*

| Article title | Global citations No. | Global citations Per year | Local citations No. | Publication year | Journal | Type of paper |
|---|---|---|---|---|---|---|
| Demand side management: demand response, intelligent energy systems, and smart loads (Palensky and Dietrich 2011) | 1,518 | 169 | 44 | 2011 | IEEE Transactions on Industrial Informatics | Review paper |
| Understanding carbon lock-in (Unruh 2000) | 1,191 | 60 | 66 | 2000 | Energy Policy | Position paper |
| Contrasting the capabilities of building energy performance simulation programs (Crawley et al. 2008) | 849 | 71 | 18 | 2008 | Building and Environment | Review paper |
| The future of anaerobic digestion and biogas utilisation (Holm-Nielsen et al. 2009) | 820 | 75 | 7 | 2009 | Bioresource Technology | Position paper |
| Providing all global energy with wind, water, and solar power, part I: technologies, energy resources, quantities and areas of infrastructure, and materials (Jacobson and Delucchi 2011) | 766 | 85 | 44 | 2011 | Energy Policy | Research/review paper |
| Calculation of radiative fluxes from the surface to top of atmosphere based on ISCCP and other global data sets: refinements of the radiative transfer model and the input data (Zhang et al. 2004) | 751 | 47 | 8 | 2004 | Journal of Geophysical Research D: Atmospheres | Research paper |
| RCP4.5: a pathway for stabilisation of radiative forcing by 2100 (Thomson et al. 2011) | 693 | 77 | 10 | 2011 | Climatic Change | Research paper |
| Renewable energy strategies for sustainable development (Lund 2007) | 680 | 52 | 71 | 2007 | Energy | Research paper |
| Applications of artificial neural-networks for energy systems (Kalogirou 2000) | 636 | 32 | 16 | 2000 | Applied Energy | Review paper |
| Energy system analysis of 100% renewable energy systems-the case of Denmark in years 2030 and 2050 (Lund and Mathiesen 2009) | 626 | 57 | 65 | 2009 | Energy | Research paper |
| Integration of renewable energy into the transport and electricity sectors through V2G (Lund and Kempton 2008) | 616 | 51 | 76 | 2008 | Energy Policy | Research paper |
| Power management of a stand-alone wind/photovoltaic/fuel cell energy system (Wang and Nehrir 2008) | 561 | 47 | 37 | 2008 | IEEE Transactions on Energy Conversion | Research paper |
| The politics and policy of energy system transformation - explaining the German diffusion of renewable energy technology (Jacobsson and Lauber 2006) | 557 | 40 | 5 | 2006 | Energy Policy | Review paper |
| Energy- and greenhouse gas-based LCA of biofuel and bioenergy systems: key issues, ranges and recommendations (Cherubini et al. 2009) | 552 | 50 | 23 | 2009 | Resources, Conservation and Recycling | Review paper |
| The diffusion of renewable energy technology: an analytical framework and key issues for research (Jacobsson and Johnson 2000) | 526 | 26 | 19 | 2000 | Energy Policy | Review paper |
| The role of district heating in future renewable energy systems (Lund et al. 2010) | 503 | 50 | 112 | 2010 | Energy | Research paper |
| A novel optimisation sizing model for hybrid solar-wind power generation system (Yang et al. 2007) | 481 | 37 | 69 | 2007 | Solar Energy | Research paper |
| [4]Entropy generation in steady MHD flow due to a rotating porous disk in a nanofluid (Rashidi et al. 2013) | 472 | 67 | 2 | 2013 | International Journal of Heat And Mass Transfer | Research paper |
| Towards the hydrogen economy? (Marbán and Valdés-Solís 2007) | 460 | 35 | 1 | 2007 | International Journal of Hydrogen Energy | Position/review paper |
| Electric, hybrid, and fuel-cell vehicles: architectures and modeling (Chan et al. 2010) | 452 | 45 | 1 | 2010 | IEEE Transactions on Vehicular Technology | Review paper |
| The first step towards a 100% renewable energy-system for Ireland (Connolly et al. 2011) | 249 | 28 | 93 | 2011 | Applied Energy | Research paper |
| Large-scale integration of wind power into different energy systems (Lund 2005) | 378 | 25 | 86 | 2005 | Energy | Research paper |
| Heat Roadmap Europe: Combining district heating with heat savings to decarbonise the EU energy system (Connolly et al. 2014) | 358 | 60 | 86 | 2014 | Energy Policy | Research paper |
| *100% Renewable energy systems, climate mitigation and economic growth (Mathiesen et al. 2011)* | 374 | 42 | 84 | 2011 | Applied Energy | Research paper |
| *Pre-feasibility study of stand-alone hybrid energy systems for applications in Newfoundland (Khan and Iqbal 2005)* | 249 | 23 | 68 | 2005 | Renewable Energy | Research paper |
| *A MILP model for integrated plan and evaluation of distributed energy systems (Ren and Gao 2010)* | 378 | 28 | 68 | 2010 | Applied Energy | Research paper |

It can be seen in Table 7 that 11 out of 26 papers were review and/or position papers, and the first four of the most cited papers were all review papers (Palensky and Dietrich 2011; Unruh 2000; Holm-Nielsen et al. 2009; Crawley et al. 2008). The most influential research papers dealt with 100% renewable energy systems, hybrid microgrids, distributed systems, hydrogen economy and climate modelling. No less than six papers in Table 7 carried out for the case of Denmark – five of them dealing with 100% renewable energy systems (Lund and Kempton 2008; Lund 2007; Lund and Mathiesen 2009; Lund et al. 2010; Lund 2005; Mathiesen et al. 2011). Furthermore, it is interesting to note that none of the papers entering the list dealt with fossil fuel systems nor with nuclear energy. It seems that the research papers entering the list were first movers within specific sub-fields, such as simulation of national 100% renewable energy systems, or renewable hybrid energy systems. The three earliest publications from the year 2000 dealt with topics such as carbon lock-in to expand the debate about the energy transition (Unruh 2000), the opportunities of artificial neural-networks (Kalogirou 2000), and the promising diffusion of competitive renewable energy technology while also mentioning the associated uncertainty (Jacobsson and Johnson 2000). The newest publication from the year 2014 demonstrated the usefulness of a large-scale renewable district heat strategy to reduce greenhouse gases in Europe (Connolly et al. 2014). Six of the papers dealing with 100% renewable energy systems were carried out in EnergyPLAN modelling (Connolly et al. 2014; Lund 2007; Lund and Mathiesen 2009; Lund et al. 2010; Lund 2005; Mathiesen et al. 2011), out of 26 papers considered in total. Two papers from the list dealt with climate modelling – more specifically with modelling of Earth's surface radiation (Zhang et al. 2004; Thomson et al. 2011). Lund H was a co-author of no less than eight influential papers (Connolly et al. 2014; Lund and Kempton 2008; Lund 2007; Lund and Mathiesen 2009; Lund et al. 2010; Lund 2005; Mathiesen et al. 2011; Connolly et al. 2011). The list of the most influential papers further showed the importance of the review papers in the field, which provide a concise and clear overview of different topics, summarising many relevant papers from the respective subfield.

Studies with the highest annual citation rate since 2015 can be seen in Table 8. Somewhat surprisingly, the papers from the more recent years were more represented, as 7 out of 10 papers were published in 2018 and 2019 (Zhang et al. 2004; Mengelkamp et al. 2018; Zhao et al. 2018; Guan et al. 2018; Grubler et al. 2018; Acar and Dincer 2019; Burke and Stephens 2018). The publications are spread among many different journals. Three papers more tightly related to material science entered the list, which should have been excluded by the keyword filter (cf. Section 2). Out of the remaining seven papers, hydrogen, and more generally power-to-gas, are the focus of three papers, showing that the power-to-gas technologies came to prominence in the energy systems analysis community recently. Furthermore, five of the papers were review papers, confirming the importance of review papers in the quickly developing field. Finally, one paper described the possibility of using blockchain for community microgrids leads the list, clearly denoting that blockchain technology has gained momentum in the field.

---
[4] Upon the closer examination of the paper, it was concluded that the paper does not belong to the Energy system analysis field. However, as they were captured by the used keywords, we did not remove them in order not to alternate the used methods artificially.

*Table 8: Articles among the scientific contributions on energy system analysis, which have the highest annual citation rate since 2015. The publications are ordered based on global citations per year.*

| Article title | Global citations | | Publication year | Journal |
|---|---|---|---|---|
| | No. | Per year | | |
| *Designing microgrid energy markets: a case study: the Brooklyn microgrid (Mengelkamp et al. 2018)* | 268 | 134 | 2018 | Applied Energy |
| *Future cost and performance of water electrolysis: an expert elicitation study (Schmidt et al. 2017)* | 206 | 69 | 2017 | International Journal of Hydrogen Energy |
| *[4] MoSe2 nanosheets perpendicularly grown on graphene with Mo-C bonding for sodium-ion capacitors (Zhao et al. 2018)* | 126 | 63 | 2018 | Nano Energy |
| *Power to gas: technological overview, systems analysis and economic assessment for a case study in Germany (Schiebahn et al. 2015)* | 314 | 63 | 2015 | International Journal of Hydrogen Energy |
| *[4] Progress in enhancement of $CO_2$ absorption by nanofluids: a mini review of mechanisms and current status (Zhang et al. 2018)* | 115 | 58 | 2018 | Renewable Energy |
| *[4] Hollow Mo-doped CoP nanoarrays for efficient overall water splitting (Guan et al. 2018)* | 113 | 57 | 2018 | Nano Energy |
| *A low energy demand scenario for meeting the 1.5°C target and sustainable development goals without negative emission technologies (Grubler et al. 2018)* | 106 | 53 | 2018 | Nature Energy |
| *Review and evaluation of hydrogen production options for better environment (Acar and Dincer 2019)* | 53 | 53 | 2019 | Journal of Cleaner Production |
| *How long will it take? Conceptualising the temporal dynamics of energy transitions (Sovacool 2016)* | 209 | 52 | 2016 | Energy Research and Social Science |
| *Political power and renewable energy futures: a critical review (Burke and Stephens 2018)* | 98 | 49 | 2018 | Energy Research and Social Science |

### 3.4. Keywords and themes

The most relevant keywords in the ESA literature are presented in Figure 6. Keywords with the highest mean publication year were *Energy transition*, *Multi-objective optimisation* and *Microgrid*, while the keyword with the most appearances was *Renewable energy*. Only one keyword (*Modelling*) did not have a rising trend in its use, which is in line with the growing number of papers published in the field in general. However, this keyword was likely replaced by more specific keywords such as *Optimisation, Multi-objective optimisation*, and *Energy system modelling,* which all had later mean publishing years. The importance of the *Modelling* keyword can be seen by the above-average mean global citations of publications with that keyword. The keyword with the highest mean global citations was *District heating*. A further look also revealed the importance of the ESA research field for a sustainable and decarbonised future system. Thereby, keywords like *Renewable energy, Solar energy, Energy efficiency, Wind energy, Renewable energy systems, Photovoltaic, Climate change, Sustainable* and *Energy transition* made up the core of the publications.

Moreover, there were some keywords with a high number of global citations that had a low number of appearances. For example, *energy system analysis, district heating, biomass* and *bioenergy* keywords had very high mean global citations indicators but a very low number of appearances. Those keywords are very broad, and thus, those could be used for review papers that appeared to be very influential (Table 7). On the other hand, keywords like *Renewable energy* and *Optimisation* had a high number of appearances but a low global mean citations indicator. This shows that they also appeared in many low impact publications that dragged the global mean citations indicator down. Furthermore, a keyword such as *Optimisation* could also be used in articles that are tackling very specific and complex problems which will often have a lower number of citations due to specificity of the subfield.

Furthermore, according to our clustering of the individual keywords, fundamental aspects and multi-generation aspects were discussed extensively in the first half of the past decade. In the past few years, grid and methodological aspects seem to play a more important role. Thereby, various renewable aspects remained a focus of research throughout this whole period. Furthermore, the more specific and descriptive keywords started appearing in the last few years, as the development of the field of ESA gained momentum. For example, *Multi-objective optimisation* instead of *Optimisation*, *Integrated energy systems* instead of *Energy systems* and *Hybrid renewable energy systems* instead of *Renewable energy systems* were used more often. The focus on the methods but also more specific keywords can indicate that the results achieved so far started to be tested concerning robustness. The highest appearances, denoted with the largest bubbles, are associated with the keywords *Optimisation*, *Renewable energy*, and *Energy systems*. All of the latter keywords had a mean occurrence in the period 2014-2015. The keywords with the latest mean occurrence are *Integrated energy systems*, *Energy transition* and *Demand response*, at least two of which highlight increased attention on sector coupling and demand-side flexibility. Thus, the general tendency shows that as the number of keywords increased in the last years, becoming more descriptive, their global average citations count became smaller in absolute numbers.

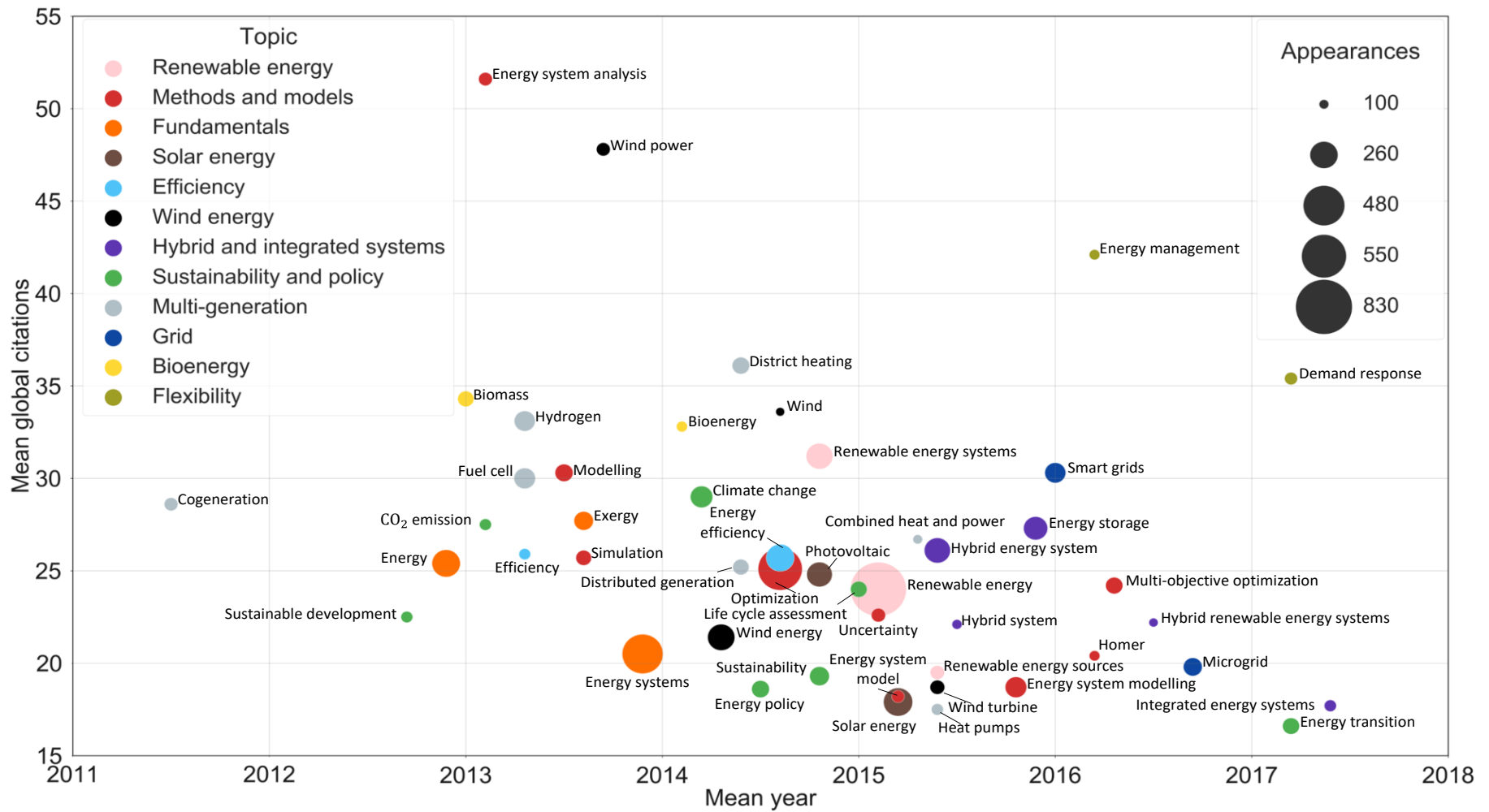

Figure 6: 50 most relevant keywords in publications on energy system analysis in terms of the total number of appearances in the period 2000-2019. The x-axis shows the mean year of publication for each keyword. The bubble size indicates the total number of appearances. The keywords are grouped according to their topic by manual classification based on the author's expert judgement. Keyword trends can be further examined in Appendix A. The mean global citations represent the global citations of the publications where the keyword is used averaged over the total number of publications where the keyword is used. For example, if two articles contain the keyword "District heating" among the articles that are analysed, one with ten and the other with 15 global citations, then this keyword would have (10+20)/2=15 mean global citations. The appearances present the number of times the keyword was used in publications.

## 4. Discussion

This review paper has highlighted several important findings. Firstly, there was neither correlation between the share of collaborative publications (the authors' institutions that do not belong to the same country) and the total number of publications nor with any of the h-, m- and g-indices. Two potential reasons can explain why the latter was observed. First, different co-authors, having different educational backgrounds could end up working in the same affiliation, as a result of the previous collaborations. The second potential reason is that the policy specifics are very different from country to country, which presents a hindrance to the potential wider collaboration between the authors. Although most of the research on collaboration focuses on the individual authors' collaborations, one of the papers focusing on extramural collaboration indeed confirmed our finding that there is no clear correlation among the two (Abramo et al. 2009). Secondly, Aalborg University has four authors among the most productive ones, and they collaborated on five of the most cited publications. Hence, it can be seen that there is a strong attraction force among the most active authors and their affiliations within the field. The latter can be a good sign if it succeeds in increasing the exchange of ideas inside the group. On the other hand, in the long run, it can also curb the visibility of scientific debate to the wider community, if the most prominent scientific questions are debated within certain affiliations.

In addition, this paper presented the resulting trends regarding countries, institutions, authors, journals and keywords. As the most recent dynamics are expected to persist in the near future, one can estimate dominant groups, locations and topics in the field. First, China has by far the highest slope of the number of publications, showing its importance for the field in the future. Second, all the most productive universities show an increasing trend in the number of publications, at least from 2015. The first three institutions with the highest slope are *North China Electric Power University, Ministry of Education China* and *the Technical University of Denmark* (in that order). All three of them find their way to the list of the most productive institutions, although none of them topped the list. Third, trends regarding the most productive authors show mixed results, as 8 out of 20 authors that entered the list show a declining trend in the number of publications. Fourth, there are three journals with a significantly steeper growing trend than the others: *Energies*, *Applied Energy* and *Energy* (in that order). Especially the journal *Energies* is dominating the rising trend, which confirms the rising trend in open-access publishing, as the *Energies* journal is completely open-access. Fifth, the keywords *Renewable Energy, Optimization* and *Solar Energy* (in that order) showed the strongest upward trend in the field. On the other hand, only one keyword from the list had a declining trend, *Modelling*. The latter shows that renewable energy systems, in general, are taking a strong footing in the field in general. Moreover, none of the keywords that could be associated with fossil fuels managed to enter the list of the 20 most frequent keywords, confirming the tendency of the field to put more emphasis on the research of renewable energy systems.

While this paper presents a far more comprehensive and deeper assessment of existing peer-reviewed papers with a focus on energy system analysis and modelling, there are existing bibliometric analyses that have covered some parts of the work carried out in this paper. In a quite broad analysis of energy-related papers from the year 2014 to 2017 of all fields without focusing on peer-reviewed publications, Angulo-Cuentas and Mongua (2018) point out that China, the USA, India, United Kingdom, and Germany are the countries with the most publications in the corresponding field. Despite the minor differences in the ranking (Figure 3), this finding was confirmed in this paper, where it was shown that the USA, China, the UK, Germany and India (in that order) are the most productive countries. The same or nearly the same countries in slightly different order also lead the ranking in more specific bibliometric analyses with a focus on peer-reviewed journal articles such as Weinand (2020a) on municipal energy system planning (China, USA, Germany, UK, Canada), Akbari et al. (2020) on sustainable technology research (USA, China, UK, India, Germany), and Zhou et al. (2018) on energy security (USA, China, UK, India, Germany). In contrast, the bibliometric analysis by Janik et al. (2020) on sustainable city landscapes lists Italy and Spain as the countries with the highest number of total publications. In this paper, Italy and Spain were placed six[th] and 9[th] in the list of the most productive countries. This discrepancy shows that some countries were more productive in specific subfields of the ESA field, while they were less productive in the ESA field in general. Furthermore, the study by Janik et al. (2020) concludes that the USA and the UK are the countries with the strongest collaboration. Since this paper showed that authors from one country mostly work with authors from the same country, this seems to be contrary to our results. Regarding the origin of the most important institutions, Angulo-Cuentas and Mongua (2018) have found that they come from China and Denmark, a finding that was confirmed in this paper. Furthermore, according to Angulo-Cuentas and Mongua (2018), the Journal of Power Sources and the Journal of Cleaner Production are the most productive journals in the respective fields. The latter is also at the top in the case of Akbari et al. (2020). In this paper, the Journal of Cleaner Production was ranked 8[th] in the list of the most productive journals (Table 6). Nevertheless, most of the productive journals listed in this paper, such as Energy, Applied Energy, Energy Policy, Energies, and Renewable Energy are also in the front ranks of further bibliometric analyses. This is especially true for the analyses of Weinand (2020a) who also states that the top journals (Energy, Applied Energy, Energy Policy, Energies, Renewable Energy) have published 37% of the retrieved articles. The ranking of Zhou et al. (2018) covers similar journals but concludes that Energy Policy is the most important journal. In contrast, the ranking of Akbari et al. (2020) lists a quite different top 5 to our presented results: Journal of Cleaner Production, Abstracts of papers of the American Chemical Society, Sustainability, Energy Policy, Clean Technologies and Environmental Policy Production. Additionally, the presented keywords and topics differ widely. This can be attributed to the different focus of the bibliometric analyses. While Weinand (2020a) states that the major focus of the publications is on renewable energies and optimisation, according to him, the top keywords are energy fuels, thermodynamics, environmental sciences, and green technologies. Moreover, Janik et al. (2020) presented the core topics with smart cities, sustainable cities, sustainable development, ICT, big data analytics, and urban sustainability, as well as Akbari et al. (2020), revealed five topic clusters with the competitive advantage of environmental innovation, development of sustainable technologies, environmental policy tools, undesirable output solutions, innovative environmental activities.

Several limitations arose in this study. First, the analysis of keywords did not include a check of the relevance of each paper itself due to the very high number of publications involved in the analysis[4]. Consequently, some of the papers entering the analysis were not always relevant to the field. Second, regarding the analysis of authors and institutions, one author could have changed affiliation one or several times throughout her career. In that case, all the publications were attributed to the first affiliation of an author having several affiliation changes. Third, some smaller countries having high indicators per capita could have been left out of the lists, as the lists themselves were based on the absolute numbers and then expressing other indicators that were presented in relative numbers. Fourth, although our Scopus search focused only on the papers classified as articles and not including review papers, Table 7 reveals that many of the papers that passed through the Scopus filter were review papers. Hence, it can be concluded that the total number of 12,182 articles included in this analysis also included some of the research papers. Fifth, this level of analysis involving a combination of bibliometric and manual analysis remains by definition at quite a high level. Whilst we were able to explore broad keywords, for example, it was not feasible to delve into the details of the individual models and methodologies. What the study loses in-depth, it makes up for in breadth.

# 5. Summary and conclusions

A comprehensive bibliometric analysis on ESA was carried out in this paper for the first time, by employing different algorithms in Matlab and R. The focus of *results* was on quantitative indicators relating to the number and type of publication outputs. Furthermore, collaboration links between institutions, authors and countries were established, as well as dynamic trends within the field.

The field of ESA has experienced exponential growth in the number of publications since at least the year 2000. The five most productive countries have 50% of the publications on ESA (representing only 25% of the world population), while the twelve most productive countries have more than 80% of the publications. The dominant institutions are even more concentrated within a small number of institutions and countries. A significant concentration of published papers within countries and institutions was also confirmed by the analysis of collaboration networks. The collaboration networks of both institutions and authors show that they dominantly collaborate within the same university, as well as within the same country. There is also is a strong link among the most successful journals, authors and institutions. The *Energy* journal has had the most publications in the field, and its editor-in-chief is Lund H. Lund H is the author with most of the publications in the field, as well as the author with most of the highly cited publications in the field. In terms of the dynamics within the field in the past decade, recent years have seen more focus on topics related to flexibility and hybrid/integrated energy systems alongside a decline in individual technologies.

The advantage of the approach in terms of a wide scope is also its disadvantage in terms of depth. On the one hand, it was possible to identify broad themes and publication trends in the ESA field during the past few decades. On the other hand, it was not feasible to explore specific methods and models in order to have a more critical analysis of the research content. This is an inherent limitation of such bibliometric approaches which tend to focus more on outputs themselves rather than their content.

Nevertheless, by taking stock in this way, the study still has several implications for the field. It provides a holistic overview of two decades' research output and enables interested readers to obtain a comprehensive overview of the key trends in the field. It could also provide a solid foundation for interested scholars to expand their collaboration networks in order to involve a more diverse group of collaborators, potentially increasing the relevance of analyses in the field. Finally, the results presented in this study can support researchers new to the field in a deeper understanding of trends and key authors (researchers, institutions, countries) internationally.

## Acknowledgements

Dominik Franjo Dominković was funded by the CITIES project nr. DSF1305-00027B funded by the Danish Innovationsfonden. Fabian Scheller kindly acknowledges the financial support of the European Union's Horizon 2020 research and innovation programme under the Marie Sklodowska-Curie grant agreement no. 713683 (COFUNDfellowsDTU).

## Appendix A

*Table A1: The 25 most relevant keywords in publications on energy system analysis divided into categories. The percentage of values refer to a total of 12,182 publications.*

| Keywords | Appearances | | Trend | Mean year | Mean global citations |
|---|---|---|---|---|---|
| | No. | % | | | |
| Renewable energy | 828 | 6.8 | ↑ | 2015.1 | 24.0 |
| Optimization | 547 | 4.5 | ↑ | 2014.6 | 25.1 |
| Energy systems | 480 | 3.9 | ↗ | 2013.9 | 20.5 |
| Solar energy | 286 | 2.3 | ↗ | 2015.2 | 17.9 |
| Energy | 274 | 2.2 | ↗ | 2012.9 | 25.4 |
| Energy efficiency | 271 | 2.2 | ↗ | 2014.6 | 25.7 |
| Wind energy | 260 | 2.1 | ↗ | 2014.3 | 21.4 |
| Renewable energy systems | 253 | 2.1 | ↗ | 2014.8 | 31.2 |
| Hybrid energy system | 245 | 2.0 | ↗ | 2015.4 | 26.1 |
| Photovoltaic | 239 | 2.0 | ↗ | 2014.8 | 24.8 |
| Energy storage | 222 | 1.8 | ↗ | 2015.9 | 27.3 |
| Climate change | 203 | 1.7 | ↗ | 2014.2 | 29.0 |
| Energy system modelling | 194 | 1.6 | ↗ | 2015.8 | 18.7 |
| Fuel cell | 193 | 1.6 | ↗ | 2013.3 | 30.0 |
| Hydrogen | 188 | 1.5 | ↗ | 2013.3 | 33.1 |
| Smart grids | 188 | 1.5 | ↗ | 2016.0 | 30.3 |
| Sustainability | 176 | 1.4 | ↗ | 2014.8 | 19.3 |
| Exergy | 173 | 1.4 | ↗ | 2013.6 | 27.7 |
| Microgrid | 168 | 1.4 | ↗ | 2016.7 | 19.8 |
| Modelling | 161 | 1.3 | ↓ | 2013.5 | 30.3 |
| Energy policy | 159 | 1.3 | ↗ | 2014.5 | 18.6 |
| District heating | 156 | 1.3 | ↗ | 2014.4 | 36.1 |
| Multi-objective optimization | 154 | 1.3 | ↗ | 2016.3 | 24.2 |
| Energy transition | 152 | 1.2 | ↗ | 2017.2 | 16.6 |
| Life cycle assessment | 149 | 1.2 | ↗ | 2015.0 | 24.0 |

The matrix of correlations between different keywords can be seen in Figure A1. The most correlated keyword pairs are *Exergy-Energy*, *Optimisation-Renewable energy*, *Wind energy-Solar energy*, *Energy storage-Renewable energy*, and *Energy systems-Optimization*. In this context, keywords with a joint appearance of over 30 can be seen as important. Due to the high correlations in terms of four pairs, the keyword *Renewable energy* plays a particularly important role in the analysed body of ESA literature. This underlines once again the importance of ESA in tackling future transformative challenges.

| | A | B | C | D | E | F | G | H | I | J | K | L | M | N | O | P | Q | R | S | T | U | V | W | X | Y | |
|---|---|---|---|---|---|---|---|---|---|---|---|---|---|---|---|---|---|---|---|---|---|---|---|---|---|---|---|
| A | 203 | 20 | 0 | 0 | 16 | 3 | 0 | 11 | 3 | 7 | 0 | 1 | 8 | 8 | 1 | 5 | 3 | 0 | 1 | 1 | 12 | 0 | 1 | 0 | 6 | Climate change (A) |
| B | | 480 | 9 | 1 | 2 | 7 | 6 | 14 | 8 | 2 | 4 | 7 | 41 | 6 | 18 | 10 | 7 | 1 | 5 | 0 | 21 | 8 | 6 | 3 | 22 | Energy systems (B) |
| C | | | 156 | 0 | 2 | 1 | 0 | 2 | 8 | 10 | 1 | 2 | 17 | 2 | 3 | 0 | 0 | 0 | 3 | 10 | 1 | 3 | 0 | 0 | 1 | District heating (C) |
| D | | | | 245 | 2 | 12 | 18 | 0 | 5 | 1 | 0 | 12 | 24 | 3 | 3 | 0 | 5 | 12 | 24 | 4 | 38 | 11 | 2 | 13 | 1 | Hybrid energy system (D) |
| E | | | | | 274 | 3 | 8 | 0 | 8 | 0 | 65 | 7 | 16 | 6 | 0 | 0 | 11 | 2 | 1 | 1 | 13 | 2 | 2 | 11 | 16 | Energy (E) |
| F | | | | | | 220 | 4 | 3 | 2 | 9 | 21 | 1 | 7 | 1 | 0 | 0 | 14 | 12 | 9 | 17 | 43 | 2 | 9 | 9 | 3 | Energy storage (F) |
| G | | | | | | | 193 | 1 | 6 | 0 | 5 | 13 | 9 | 3 | 7 | 0 | 31 | 5 | 13 | 4 | 20 | 4 | 6 | 10 | 4 | Fuel cell (G) |
| H | | | | | | | | 159 | 6 | 9 | 0 | 2 | 2 | 2 | 3 | 11 | 3 | 0 | 1 | 3 | 21 | 0 | 2 | 1 | 5 | Energy policy (H) |
| I | | | | | | | | | 265 | 4 | 7 | 2 | 15 | 4 | 4 | 3 | 6 | 3 | 1 | 5 | 26 | 1 | 9 | 2 | 2 | Energy efficiency (I) |
| J | | | | | | | | | | 194 | 0 | 1 | 9 | 9 | 0 | 6 | 3 | 0 | 3 | 2 | 17 | 0 | 0 | 2 | 9 | Energy system modelling (J) |
| K | | | | | | | | | | | 169 | 3 | 21 | 3 | 0 | 0 | 10 | 0 | 1 | 0 | 8 | 1 | 0 | 18 | 9 | Exergy (K) |
| L | | | | | | | | | | | | 239 | 17 | 3 | 1 | 4 | 6 | 7 | 11 | 4 | 30 | 3 | 4 | 16 | | Photovoltaic (L) |
| M | | | | | | | | | | | | | 547 | 3 | 15 | 1 | 8 | 16 | 9 | 18 | 60 | 1 | 7 | | 12 | Optimization (M) |
| N | | | | | | | | | | | | | | 149 | 0 | 0 | 4 | 2 | 2 | 2 | 10 | 4 | 0 | 6 | 7 | Life cycle assessment (N) |
| O | | | | | | | | | | | | | | | 161 | 1 | 6 | 2 | 4 | 7 | 15 | 1 | 2 | 4 | 5 | Modelling (O) |
| P | | | | | | | | | | | | | | | | 152 | 0 | 2 | 1 | 0 | 11 | 0 | 2 | 0 | 4 | Energy transition (P) |
| Q | | | | | | | | | | | | | | | | | 188 | 1 | 5 | 2 | 19 | 1 | 3 | 9 | 11 | Hydrogen (Q) |
| R | | | | | | | | | | | | | | | | | | 168 | 3 | 3 | 23 | 1 | 11 | 1 | 0 | Microgrid (R) |
| S | | | | | | | | | | | | | | | | | | | 260 | 12 | 29 | 3 | 3 | 57 | 10 | Wind energy (S) |
| T | | | | | | | | | | | | | | | | | | | | 253 | 3 | 5 | 6 | 10 | 18 | Renewable energy systems (T) |
| U | | | | | | | | | | | | | | | | | | | | | 828 | 7 | 16 | 34 | 18 | Renewable energy (U) |
| V | | | | | | | | | | | | | | | | | | | | | | 154 | 2 | 5 | 1 | Multi-objective optimization (V) |
| W | | | | | | | | | | | | | | | | | | | | | | | 184 | 2 | 4 | Smart grids (W) |
| X | | | | | | | | | | | | | | | | | | | | | | | | 286 | 2 | Solar energy (X) |
| Y | | | | | | | | | | | | | | | | | | | | | | | | | 176 | Sustainability (Y) |

*Figure A1: Correlation matrix of the 25 most relevant keywords in the research field of energy system analysis. The numbers indicate how often the keywords appear together in publications. The darker the fields in the matrix are coloured green, the more often these keywords appear together.*

# References

Abramo, Giovanni; D'Angelo, Ciriaco Andrea; Di Costa, Flavia (2009): Research collaboration and productivity: is there correlation? In *High Educ* 57 (2), pp. 155–171. DOI: 10.1007/s10734-008-9139-z.

Acar, Canan; Dincer, Ibrahim (2019): Review and evaluation of hydrogen production options for better environment. In *Journal of Cleaner Production* 218, pp. 835–849. DOI: 10.1016/j.jclepro.2019.02.046.

Akbari, Morteza; Khodayari, Maryam; Danesh, Mozhgan; Davari, Ali; Padash, Hamid (2020): A bibliometric study of sustainable technology research. In *Cogent Business & Management* 7 (1). DOI: 10.1080/23311975.2020.1751906.

ANGULO-CUENTAS, GERARDO; MONGUA, SILVIA VALENZUELA (2018): Current in the energy field 2014-2017: A bibliometric analysis of scientific trends. International Association for Management of Technology (IAM0T). Available online at https://www2.aston.ac.uk/aston-business-school/documents/IAMOT2018_paper_170.pdf.

Aria, Massimo; Cuccurullo, Corrado (2017): bibliometrix. An R-tool for comprehensive science mapping analysis. In *Journal of Informetrics* 11 (4), pp. 959–975. DOI: 10.1016/j.joi.2017.08.007.

Baños, R.; Manzano-Agugliaro, F.; Montoya, F. G.; Gil, C.; Alcayde, A.; Gómez, J. (2011): Optimization methods applied to renewable and sustainable energy: A review. In *Renewable and Sustainable Energy Reviews* 15 (4), pp. 1753–1766. DOI: 10.1016/j.rser.2010.12.008.

Boustead, I.; Hancock, G. F. (1979): Handbook of industrial energy analysis. Available online at https://www.osti.gov/etdeweb/biblio/6633791.

Burke, Matthew J.; Stephens, Jennie C. (2018): Political power and renewable energy futures: A critical review. In *Energy Research & Social Science* 35, pp. 78–93. DOI: 10.1016/j.erss.2017.10.018.

Chan, C. C.; Bouscayrol, A.; Chen, K. (2010): Electric, Hybrid, and Fuel-Cell Vehicles: Architectures and Modeling. In *IEEE Trans. Veh. Technol.* 59 (2), pp. 589–598. DOI: 10.1109/TVT.2009.2033605.

Chapman, P. F.; Leach, G.; Slesser, M. (1974): 2. The energy cost of fuels. In *Energy Policy* 2 (3), pp. 231–243. DOI: 10.1016/0301-4215(74)90048-2.

Chapman, Peter F. (1975): Energy analysis of nuclear power stations. In *Energy Policy* 3 (4), pp. 285–298. DOI: 10.1016/0301-4215(75)90037-3.

Cherubini, Francesco; Bird, Neil D.; Cowie, Annette; Jungmeier, Gerfried; Schlamadinger, Bernhard; Woess-Gallasch, Susanne (2009): Energy- and greenhouse gas-based LCA of biofuel and bioenergy systems: Key issues, ranges and recommendations. In *Resources, Conservation and Recycling* 53 (8), pp. 434–447. DOI: 10.1016/j.resconrec.2009.03.013.

Connolly, D.; Lund, H.; Mathiesen, B. V.; Leahy, M. (2011): The first step towards a 100% renewable energy-system for Ireland. In *Applied Energy* 88 (2), pp. 502–507. DOI: 10.1016/j.apenergy.2010.03.006.

Connolly, D.; Lund, H.; Mathiesen, B. V.; Werner, S.; Möller, B.; Persson, U. et al. (2014): Heat Roadmap Europe: Combining district heating with heat savings to decarbonise the EU energy system. In *Energy Policy* 65, pp. 475–489. DOI: 10.1016/j.enpol.2013.10.035.

Crawley, Drury B.; Hand, Jon W.; Kummert, Michaël; Griffith, Brent T. (2008): Contrasting the capabilities of building energy performance simulation programs. In *Building and Environment* 43 (4), pp. 661–673. DOI: 10.1016/j.buildenv.2006.10.027.

DeCarolis, Joseph; Daly, Hannah; Dodds, Paul; Keppo, Ilkka; Li, Francis; McDowall, Will et al. (2017): Formalizing best practice for energy system optimization modelling. In *Applied Energy* 194, pp. 184–198. DOI: 10.1016/j.apenergy.2017.03.001.

Dincer, Ibrahim; Acar, Canan (2015): A review on clean energy solutions for better sustainability. In *Int. J. Energy Res.* 39 (5), pp. 585–606. DOI: 10.1002/er.3329.

Egghe, Leo (2006): Theory and practise of the g-index. In *Scientometrics* 69 (1), pp. 131–152. DOI: 10.1007/s11192-006-0144-7.

Elsevier (2020): Scopus: Document search. Available online at https://www.scopus.com/search/form.uri?display=basic.

Evans, Annette; Strezov, Vladimir; Evans, Tim J. (2012): Assessment of utility energy storage options for increased renewable energy penetration. In *Renewable and Sustainable Energy Reviews* 16 (6), pp. 4141–4147. DOI: 10.1016/j.rser.2012.03.048.

Grubler, Arnulf; Wilson, Charlie; Bento, Nuno; Boza-Kiss, Benigna; Krey, Volker; McCollum, David L. et al. (2018): A low energy demand scenario for meeting the 1.5 °C target and sustainable development goals without negative emission technologies. In *Nat Energy* 3 (6), pp. 515–527. DOI: 10.1038/s41560-018-0172-6.

Guan, Cao; Xiao, Wen; Wu, Haijun; Liu, Ximeng; Zang, Wenjie; Zhang, Hong et al. (2018): Hollow Mo-doped CoP nanoarrays for efficient overall water splitting. In *Nano Energy* 48, pp. 73–80. DOI: 10.1016/j.nanoen.2018.03.034.

Hall, Lisa M. H.; Buckley, Alastair R. (2016): A review of energy systems models in the UK: Prevalent usage and categorisation. In *Applied Energy* 169, pp. 607–628. DOI: 10.1016/j.apenergy.2016.02.044.

Hirsch, J. E. (2005): An index to quantify an individual's scientific research output. In *Proceedings of the National Academy of Sciences of the United States of America* 102 (46), pp. 16569–16572. DOI: 10.1073/pnas.0507655102.

Holm-Nielsen, J. B.; Al Seadi, T.; Oleskowicz-Popiel, P. (2009): The future of anaerobic digestion and biogas utilization. In *Bioresource technology* 100 (22), pp. 5478–5484. DOI: 10.1016/j.biortech.2008.12.046.

Jacobson, Mark Z.; Delucchi, Mark A. (2011): Providing all global energy with wind, water, and solar power, Part I: Technologies, energy resources, quantities and areas of infrastructure, and materials. In *Energy Policy* 39 (3), pp. 1154–1169. DOI: 10.1016/j.enpol.2010.11.040.

Jacobsson, Staffan; Johnson, Anna (2000): The diffusion of renewable energy technology: an analytical framework and key issues for research. In *Energy Policy* 28 (9), pp. 625–640. DOI: 10.1016/S0301-4215(00)00041-0.

Jacobsson, Staffan; Lauber, Volkmar (2006): The politics and policy of energy system transformation—explaining the German diffusion of renewable energy technology. In *Energy Policy* 34 (3), pp. 256–276. DOI: 10.1016/j.enpol.2004.08.029.

Janik, Agnieszka; Ryszko, Adam; Szafraniec, Marek (2020): Scientific Landscape of Smart and Sustainable Cities Literature: A Bibliometric Analysis. In *Sustainability* 12 (3), p. 779. DOI: 10.3390/su12030779.

Jenkins, Kirsten; McCauley, Darren; Heffron, Raphael; Stephan, Hannes; Rehner, Robert (2016): Energy justice: A conceptual review. In *Energy Research & Social Science* 11, pp. 174–182. DOI: 10.1016/j.erss.2015.10.004.

Kalogirou, Soteris A. (2000): Applications of artificial neural-networks for energy systems. In *Applied Energy* 67 (1-2), pp. 17–35. DOI: 10.1016/S0306-2619(00)00005-2.

Khan, M. J.; Iqbal, M. T. (2005): Pre-feasibility study of stand-alone hybrid energy systems for applications in Newfoundland. In *Renewable Energy* 30 (6), pp. 835–854. DOI: 10.1016/j.renene.2004.09.001.

Khare, Vikas; Nema, Savita; Baredar, Prashant (2016): Solar–wind hybrid renewable energy system: A review. In *Renewable and Sustainable Energy Reviews* 58, pp. 23–33. DOI: 10.1016/j.rser.2015.12.223.

Kondziella, Hendrik; Bruckner, Thomas (2016): Flexibility requirements of renewable energy based electricity systems – a review of research results and methodologies. In *Renewable and Sustainable Energy Reviews* 53, pp. 10–22. DOI: 10.1016/j.rser.2015.07.199.

Kumar, Abhishek; Sah, Bikash; Singh, Arvind R.; Deng, Yan; He, Xiangning; Kumar, Praveen; Bansal, R. C. (2017): A review of multi criteria decision making (MCDM) towards sustainable renewable energy development. In *Renewable and Sustainable Energy Reviews* 69, pp. 596–609. DOI: 10.1016/j.rser.2016.11.191.

Levenshtein, V. I. (1966): Binary codes capable of correcting deletions, insertions, and reversals. In *Cybernetics and Control Theory* 10 (8).

Lund, H.; Mathiesen, B. V. (2009): Energy system analysis of 100% renewable energy systems—The case of Denmark in years 2030 and 2050. In *Energy* 34 (5), pp. 524–531. DOI: 10.1016/j.energy.2008.04.003.

Lund, H.; Möller, B.; Mathiesen, B. V.; Dyrelund, A. (2010): The role of district heating in future renewable energy systems. In *Energy* 35 (3), pp. 1381–1390. DOI: 10.1016/j.energy.2009.11.023.

Lund, Henrik (2005): Large-scale integration of wind power into different energy systems. In *Energy* 30 (13), pp. 2402–2412. DOI: 10.1016/j.energy.2004.11.001.

Lund, Henrik (2007): Renewable energy strategies for sustainable development. In *Energy* 32 (6), pp. 912–919. DOI: 10.1016/j.energy.2006.10.017.

Lund, Henrik; Kempton, Willett (2008): Integration of renewable energy into the transport and electricity sectors through V2G. In *Energy Policy* 36 (9), pp. 3578–3587. DOI: 10.1016/j.enpol.2008.06.007.

Man, Jonathan P.; Weinkauf, Justin G.; Tsang, Monica; Sin, Don D. (2004): Why do some countries publish more than others? An international comparison of research funding, English proficiency and publication output in highly ranked general medical journals. In *European journal of epidemiology* 19 (8), pp. 811–817. DOI: 10.1023/B:EJEP.0000036571.00320.b8.

Mancarella, Pierluigi (2014): MES (multi-energy systems): An overview of concepts and evaluation models. In *Energy* 65, pp. 1–17. DOI: 10.1016/j.energy.2013.10.041.

Marbán, Gregorio; Valdés-Solís, Teresa (2007): Towards the hydrogen economy? In *International Journal of Hydrogen Energy* 32 (12), pp. 1625–1637. DOI: 10.1016/j.ijhydene.2006.12.017.

Martín-Gamboa, Mario; Iribarren, Diego; García-Gusano, Diego; Dufour, Javier (2017): A review of life-cycle approaches coupled with data envelopment analysis within multi-criteria decision analysis for sustainability assessment of energy systems. In *Journal of Cleaner Production* 150, pp. 164–174. DOI: 10.1016/j.jclepro.2017.03.017.

Mathiesen, Brian Vad; Lund, Henrik; Karlsson, Kenneth (2011): 100% Renewable energy systems, climate mitigation and economic growth. In *Applied Energy* 88 (2), pp. 488–501. DOI: 10.1016/j.apenergy.2010.03.001.

Mengelkamp, Esther; Gärttner, Johannes; Rock, Kerstin; Kessler, Scott; Orsini, Lawrence; Weinhardt, Christof (2018): Designing microgrid energy markets. In *Applied Energy* 210, pp. 870–880. DOI: 10.1016/j.apenergy.2017.06.054.

Meo, Sultan Ayoub; Al Masri, Abeer A.; Usmani, Adnan Mahmood; Memon, Almas Naeem; Zaidi, Syed Ziauddin (2013): Impact of GDP, spending on R&D, number of universities and scientific journals on research publications among Asian countries. In *PloS one* 8 (6), e66449. DOI: 10.1371/journal.pone.0066449.

Olatomiwa, Lanre; Mekhilef, Saad; Ismail, M. S.; Moghavvemi, M. (2016): Energy management strategies in hybrid renewable energy systems: A review. In *Renewable and Sustainable Energy Reviews* 62, pp. 821–835. DOI: 10.1016/j.rser.2016.05.040.

Palensky, Peter; Dietrich, Dietmar (2011): Demand Side Management: Demand Response, Intelligent Energy Systems, and Smart Loads. In *IEEE Trans. Ind. Inf.* 7 (3), pp. 381–388. DOI: 10.1109/TII.2011.2158841.

Pfenninger, Stefan; Hawkes, Adam; Keirstead, James (2014): Energy systems modeling for twenty-first century energy challenges. In *Renewable and Sustainable Energy Reviews* 33, pp. 74–86. DOI: 10.1016/j.rser.2014.02.003.

Rashidi, M. M.; Abelman, S.; Freidooni Mehr, N. (2013): Entropy generation in steady MHD flow due to a rotating porous disk in a nanofluid. In *International Journal of Heat and Mass Transfer* 62, pp. 515–525. DOI: 10.1016/j.ijheatmasstransfer.2013.03.004.

Ren, Hongbo; Gao, Weijun (2010): A MILP model for integrated plan and evaluation of distributed energy systems. In *Applied Energy* 87 (3), pp. 1001–1014. DOI: 10.1016/j.apenergy.2009.09.023.

Scheller, Fabian; Johanning, Simon; Bruckner, Thomas (2019): A review of designing empirically grounded agent-based models of innovation diffusion: Development process, conceptual foundation and research agenda. In *Research Report No.01 (2019), Leipzig University, Institute for Infrastructure and Resources Management (IIRM)*. Available online at hdl.handle.net/10419/191981.

Schiebahn, Sebastian; Grube, Thomas; Robinius, Martin; Tietze, Vanessa; Kumar, Bhunesh; Stolten, Detlef (2015): Power to gas: Technological overview, systems analysis and economic assessment for a case study in Germany. In *International Journal of Hydrogen Energy* 40 (12), pp. 4285–4294. DOI: 10.1016/j.ijhydene.2015.01.123.

Schmidt, O.; Gambhir, A.; Staffell, I.; Hawkes, A.; Nelson, J.; Few, S. (2017): Future cost and performance of water electrolysis: An expert elicitation study. In *International Journal of Hydrogen Energy* 42 (52), pp. 30470–30492. DOI: 10.1016/j.ijhydene.2017.10.045.

Sharma, Naveen; Varun; Siddhartha (2012): Stochastic techniques used for optimization in solar systems: A review. In *Renewable and Sustainable Energy Reviews* 16 (3), pp. 1399–1411. DOI: 10.1016/j.rser.2011.11.019.

Sovacool, Benjamin K. (2016): How long will it take? Conceptualizing the temporal dynamics of energy transitions. In *Energy Research & Social Science* 13, pp. 202–215. DOI: 10.1016/j.erss.2015.12.020.

Strachan, Neil; Fais, Birgit; Daly, Hannah (2016): Reinventing the energy modelling–policy interface. In *Nat Energy* 1 (3), pp. 1–3. DOI: 10.1038/nenergy.2016.12.

Thomson, Allison M.; Calvin, Katherine V.; Smith, Steven J.; Kyle, G. Page; Volke, April; Patel, Pralit et al. (2011): RCP4.5: a pathway for stabilization of radiative forcing by 2100. In *Climatic Change* 109 (1-2), pp. 77–94. DOI: 10.1007/s10584-011-0151-4.

Unruh, Gregory C. (2000): Understanding carbon lock-in. In *Energy Policy* 28 (12), pp. 817–830. DOI: 10.1016/S0301-4215(00)00070-7.

Wang, Caisheng; Nehrir, M. H. (2008): Power Management of a Stand-Alone Wind/Photovoltaic/Fuel Cell Energy System. In *IEEE Trans. Energy Convers.* 23 (3), pp. 957–967. DOI: 10.1109/TEC.2007.914200.

Weinand, Jann Michael (2020a): Reviewing Municipal Energy System Planning in a Bibliometric Analysis: Evolution of the Research Field between 1991 and 2019. In *Energies* 13 (6), p. 1367. DOI: 10.3390/en13061367.

Weinand, Jann Michael; Sörensen, Kenneth; Segundo, Pablo San; Kleinebrahm, Max; McKenna, Russell (2020b): Research trends in combinatorial optimisation. Available online at https://arxiv.org/pdf/2012.01294.

Yang, Hongxing; Lu, Lin; Zhou, Wei (2007): A novel optimization sizing model for hybrid solar-wind power generation system. In *Solar Energy* 81 (1), pp. 76–84. DOI: 10.1016/j.solener.2006.06.010.

Zhang, Y.; Rossow, W. B.; Lacis, A. A.; Oinas, V.; Mischenko, M. I. (2004): Calculation of radiative fluxes from the surface to top of atmosphere based on ISCCP and other global data sets: Refinements of the radiative transfer model and the input data. In *J. Geophys. Res.* 109 (D19). DOI: 10.1029/2003JD004457.

Zhang, Zhien; Cai, Jianchao; Chen, Feng; Li, Hao; Zhang, Wenxiang; Qi, Wenjie (2018): Progress in enhancement of CO2 absorption by nanofluids: A mini review of mechanisms and current status. In *Renewable Energy* 118, pp. 527–535. DOI: 10.1016/j.renene.2017.11.031.

Zhao, Xu; Cai, Wei; Yang, Ying; Song, Xuedan; Neale, Zachary; Wang, Hong-En et al. (2018): MoSe2 nanosheets perpendicularly grown on graphene with Mo–C bonding for sodium-ion capacitors. In *Nano Energy* 47, pp. 224–234. DOI: 10.1016/j.nanoen.2018.03.002.

Zhou, Wei; Kou, Aiqing; Chen, Jin; Ding, Bingqing (2018): A retrospective analysis with bibliometric of energy security in 2000–2017. In *Energy Reports* 4, pp. 724–732. DOI: 10.1016/j.egyr.2018.10.012.

Aad, G.; Abbott, B.; Abdallah, J.; Abdelalim, A. A.; Abdesselam, A.; Abdinov, O.; et al. (2013): Jet energy measurement with the ATLAS detector in proton-proton collisions at at sqrt(s) = 7 TeV. In *The European Physical Journal C* 73 (3), p. 2304. DOI: 10.1140/epjc/s10052-013-2304-2.